\documentclass{ecai} 

\usepackage{latexsym}
\usepackage{amssymb}
\usepackage{amsmath}
\usepackage{amsthm}
\usepackage{mathtools}
\usepackage{multirow}
\usepackage{booktabs}
\usepackage{enumitem}
\usepackage{graphicx}
\usepackage{array}
\usepackage{color}
\usepackage{xcolor}
\usepackage[hidelinks]{hyperref}

\newcommand{\BibTeX}{B\kern-.05em{\sc i\kern-.025em b}\kern-.08em\TeX}

\begin{document}


\begin{frontmatter}


\paperid{1850} 


\title{ReactAIvate: A Deep Learning Approach to Predicting Reaction Mechanisms and Unmasking Reactivity Hotspots}


\author[A]{\fnms{Ajnabiul}~\snm{Hoque}\orcid{0000-0001-9807-3061}\footnote{Equal contribution.}}
\author[A]{\fnms{Manajit}~\snm{Das}\orcid{0000-0001-7709-8809}\footnote{Equal contribution.}}
\author[B,C]{\fnms{Mayank}~\snm{Baranwal}\thanks{Corresponding Author. Email: baranwal.mayank@tcs.com}\orcid{0000-0001-9354-2826}}
\author[A,D]{\fnms{Raghavan}~\snm{B.	Sunoj}\thanks{Corresponding Author. Email: sunoj@chem.iitb.ac.in}\orcid{0000-0002-6484-2878}} 

\address[A]{Department of Chemistry, Indian Institute of Technology Bombay, India}
\address[B]{Department of Systems \& Control Engineering, Indian Institute of Technology, India}
\address[C]{Tata Consultancy Services Research, Mumbai, India}
\address[D]{Centre for Machine Intelligence and Data Science, Indian Institute of Technology Bombay, India}



\begin{abstract}
A chemical reaction mechanism (CRM) is a sequence of molecular-level events involving bond-breaking/forming processes, generating transient intermediates along the reaction pathway as reactants transform into products. Understanding such mechanisms is crucial for designing and discovering new reactions.  One of the currently available methods to probe CRMs is quantum mechanical (QM) computations. The resource-intensive nature of QM methods and the scarcity of mechanism-based datasets motivated us to develop reliable ML models for predicting mechanisms. In this study, we created a comprehensive dataset with seven distinct classes, each representing uniquely characterized elementary steps. Subsequently, we developed an interpretable attention-based GNN that achieved near-unity and 96\% accuracy, respectively for reaction step classification and the prediction of reactive atoms in each such step, capturing interactions between the broader reaction context and local active regions. The near-perfect classification enables accurate prediction of both individual events and the entire CRM, mitigating potential drawbacks of Seq2Seq approaches, where a wrongly predicted character leads to incoherent CRM identification. In addition to interpretability, our model adeptly identifies key atom(s) even from out-of-distribution classes. This generalizabilty allows for the inclusion of new reaction types in a modular fashion, thus will be of value to experts for understanding the reactivity of new molecules.
\end{abstract}

\end{frontmatter}


\section{Introduction}\label{sec:Intro}

The reliable prediction of chemical reactions holds paramount significance in pharmaceutical and materials manufacturing, and in understanding many processes in molecular biology ~\cite{brown2016analysis,benzinger1971thermodynamics,schupp2024chemical}. A chemical reaction entails the reorganization of atoms and bonds within the initial reactants, resulting in the creation of novel molecules or compounds as the final products. To comprehend a chemical reaction, it is essential to know the underlying chemical transformations. A sequence of these transformation steps, also called elementary mechanistic steps, is generally expressed in the form of a chemical reaction mechanism (CRM). Many of these steps may involve a bond-forming/bond-breaking process characterized by the corresponding transition state. These elementary steps serve as building blocks for developing novel reactions and discerning side products. Knowledge of CRM can provide atomic-level insights into why the products are formed \cite{goldstein1996density,chin2022computational}.

\begin{figure*}[!ht]
	\begin{center}
		\begin{tabular}{c}
			\includegraphics[width=1.8\columnwidth]{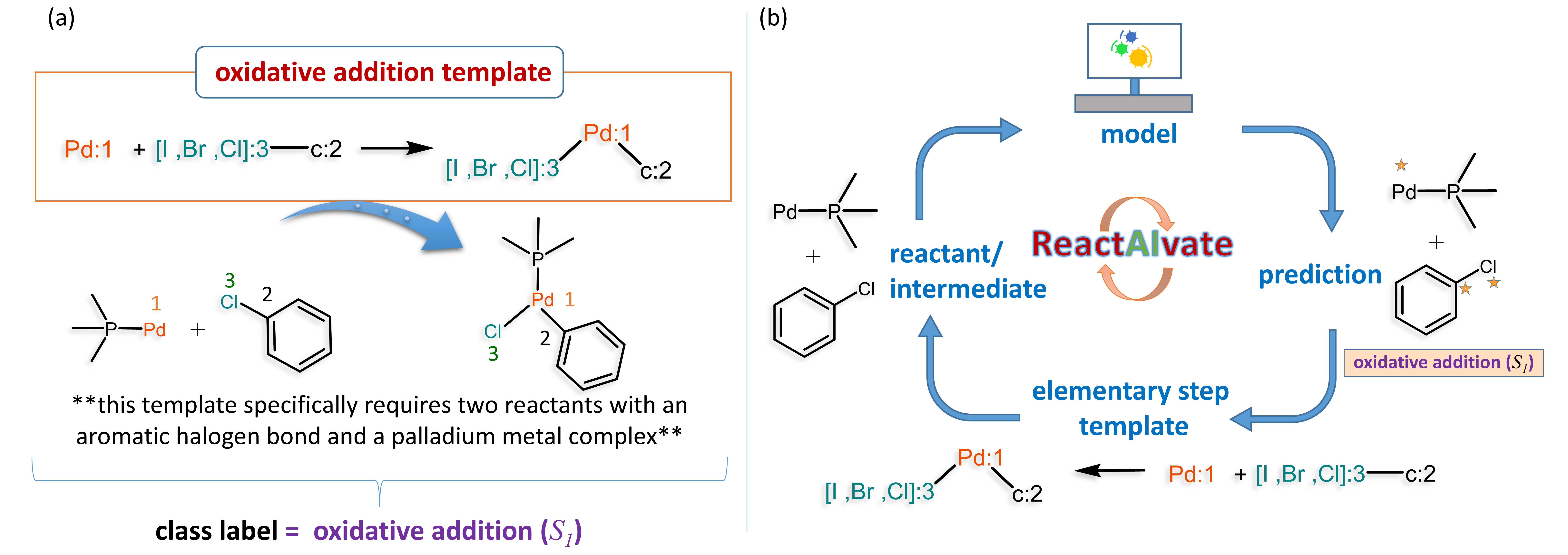} \cr
        \cr \includegraphics[width=1.8\columnwidth]{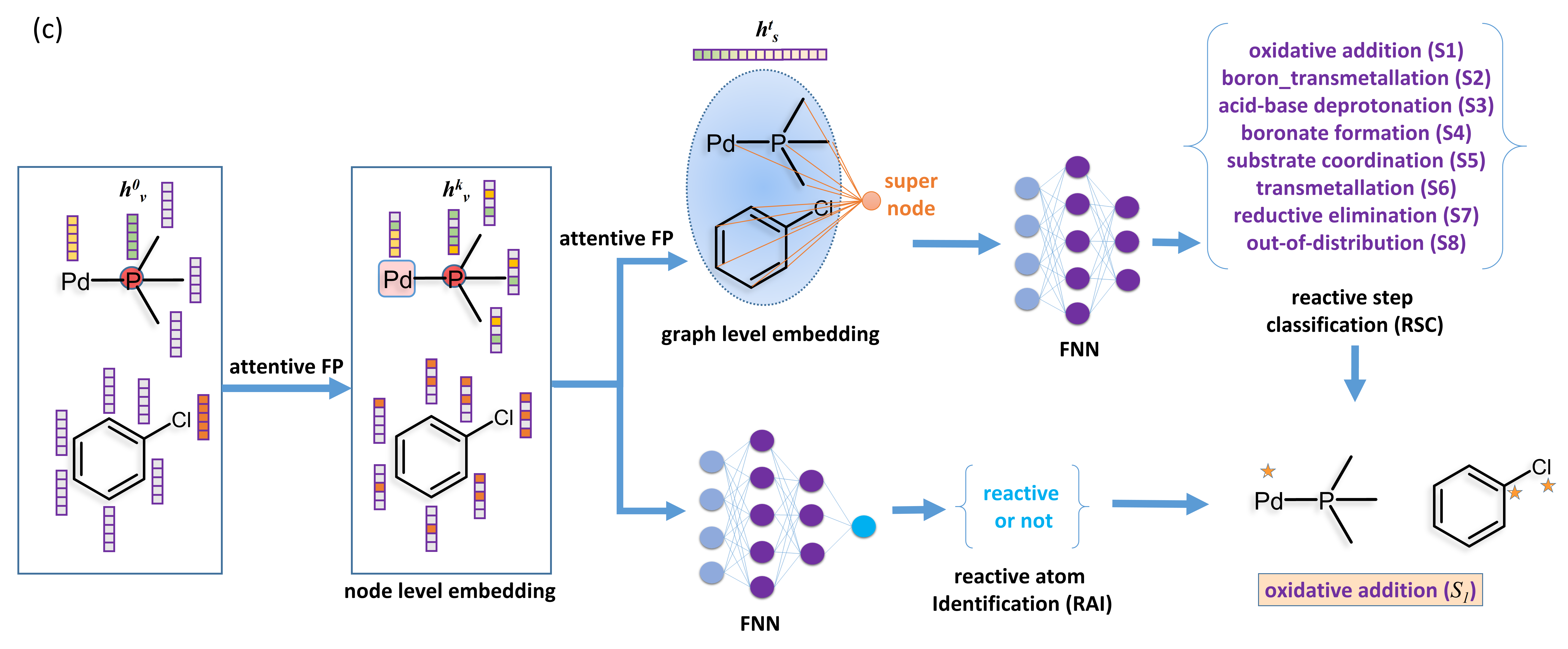}\cr
		\end{tabular}
		\caption{(a) A representative example of oxidative addition template, (b) complete workflow of our proposed ReactAIvate method, (c) process of reaction step classification and reactive atom identification using ReactAIvate}
		\label{fig1:Method}
	\end{center}
\end{figure*}

One of the ways to identify a CRM is to perform quantum mechanical (QM) calculations \cite{bhattacharya2023combinatorial,bahmanyar2003quantum}. Such methods are often computationally demanding and often require substantial human attention rendering it a time-consuming task. In recent years, numerous ML studies have focused on retrosynthesis and forward synthesis prediction, with models mostly trained on the USPTO-50k dataset \cite{schneider2016s}. While the dataset proves valuable for direct product prediction, it lacks information about elementary reaction steps, thus lacking opportunities to understand CRM ~\cite{tavakoli2023ai,chen2022generalized}. In the realm of CRM prediction, navigating the complexities of multi-step reactions and ensuring atom and charge balance presents formidable challenges. The conventional transformer-based sequence-to-sequence (Seq2Seq) models, commonly employed for sequence generation, prove inadequate in handling the intricate long-term dependencies inherent in CRM ~\cite{flam2022language}. Their limitations become glaring when even a single incorrect character introduced during inference can render the entire CRM meaningless. Beyond the conventional focus on atom and charge balance, ensuring both semantic and syntactic validity of product SMILES strings – representing molecules in a sequence-based format – highlights the necessity for a comprehensive reevaluation at the modeling level in CRM prediction. Recognizing these challenges, we have endeavored to craft an interpretable, swift, and dependable alternative for CRM prediction.

Herein, we propose an interpretable attention-based graph neural network (GNN) model for the elementary reaction step predictions, which are then used to generate the full CRM (Figure~\ref{fig1:Method}). We introduce a dataset containing 3 different families of catalytic reactions comprising 7 distinct elementary steps. In addition to the key task of identifying the elementary steps, our model is simultaneously trained to detect the reactive atoms in such steps. In the case of out-of-distribution (OOD) samples, our model points to the reactive atoms, besides classifying them into an unseen group. This can act as a guide to experts for understanding the reactivity. Below we summarize our key contributions:

\noindent \textbf{1. CRM dataset}: While there are several benchmark datasets, such as USPTO-50k, comprising single-step chemical reactions, there is currently no existing dataset on CRMs within the available literature as per our knowledge. We curate a first-of-its-kind CRM dataset containing elementary mechanistic steps for transition metal-catalyzed reactions.

\noindent \textbf{2. CRM identification via reaction step classification}: We introduce ReactAIvate, a graph-attention-based classification model that precisely identifies the necessary elementary steps for a given combination of reactants, reagents, and catalysts. Our model's accurate intermediate-step classification is pivotal for CRM identification. Notably, our approach focuses on identifying underlying reaction rules rather than generating exact product SMILES, simplifying the problem, as our experiments reveal that SMILES generation can lead to vacuous CRM predictions due to even a single character mismatch.

\noindent\textbf{3. Identification of reactive atoms/groups}: Our framework is distinctly trained to minimize a composite of two distinct loss types: (a) Graph-level loss for predicting reaction classes, and (b) Node-level loss for distinguishing reactive and non-reactive atoms. As a result, our curated database includes information on reactive atoms for each specific reaction step within a CRM. This feature is primarily introduced to offer valuable insights to domain experts.

\noindent\textbf{4. Visualizing reactive centers via attention Mechanism}: Expanding on the previous point, we demonstrate that the inclusion of node-level loss inherently compels the attention mechanism to align with reactive centers within the molecules involved. This alignment significantly enhances the visualization of reactive centers.

\noindent\textbf{5. Generalization to OOD samples}: ReactAIvate avoids overconfidence by introducing an OOD eighth class, enabling accurate identification of scenarios not covered in training data. This enhances trust in predictions and allows for the incorporation of new reaction classes when relevant data becomes available. Attention visualization on OOD samples also reveals potential reaction centers, showcasing the model's adaptability.






\section{Related Work}

Recent years have witnessed several interesting applications of machine learning in predicting molecular properties as well as their reactions \cite{wang2022multitask,ahneman2018predicting,meng2023doubly}. Intriguing ML algorithms have been developed to make complex chemical problems, such as organic synthesis, increasingly more amenable. In tasks such as forward or retrosynthesis predictions, predefined reaction templates are employed to make an intuitive connection from a set of reactants to product(s). The templates can be obtained by using data-driven approaches~\cite{segler2017neural} or through encoding by domain experts. Authors in~\cite{coley2017prediction} combined these templates and ML for product prediction and subsequent ranking of the products. Transformation rules in retrosynthesis tasks were leveraged in~\cite{segler2018planning}. Current trends suggest the use of templates for both forward and retrosynthesis analysis owing to their efficiency and interpretability.~\cite{wei2016neural,segler2017modelling,genheden2020aizynthfinder,heid2021influence}.

Interesting alternatives relying on template-free methods have emerged in very recent times. Transformer-based Seq2Seq generative model has been adopted for forward and retrosynthesis predictions using SMILES as the molecular representation~\cite{jin2017predicting,schwaller2019molecular,lin2020automatic}. Various models have been proposed that differ in molecular representation and/or model architecture~\cite{ucak2022retrosynthetic,zheng2019predicting,born2021trends,kim2021valid,wan2022retroformer}. For instance, authors in~\cite{wan2022retroformer} included both molecular graph and SMILES representations for retrosynthesis. Although the template-free methods have gained recent attention, they are known to suffer from the generation of invalid SMILES. During translation, a single character addition or a missing one can render the generated reaction invalid.

We note that the current literature although focuses on direct prediction of products and/or retrosynthesis, there are very limited efforts for CRM predictions \cite{sakai2023learning}. In~\cite{fooshee2018deep}, authors utilized deep learning on a rather limited, private dataset of elementary reactions to identify the probable electron sources and sinks and subsequent ranking of these combinations. More recently, Authors in~\cite{bradshaw2018generative} proposed ELECTRO, a graph-based generative model for electron paths movement mimicking `arrow-pushing' diagrams, but did not consider elementary reaction steps due to the absence of such details in the USPTO-50k dataset. Consequently, ELECTRO is not suitable to multi-step reactions.

        








\section{Preliminaries and Proposed Method}

\subsection{Background}

\textbf{Chemical reaction.} A chemical reaction is a process in which one set of substances (reactants) transforms into another set of substances (products). 

\noindent\textbf{Elementary step.} The elementary steps in a chemical reaction typically consists of bond breaking or bond forming involving the reactants/intermediates, where an electron rich source gets attached to an electron poor sink. The movement of electrons is usually denoted using 'arrow-pushing diagrams' where the arrow direction is from the electron source to the sink. 

\noindent\textbf{Chemical reaction mechanism (CRM).} A sequence of elementary steps that describe the transformation of reactant $\mathbf{R}$ to product $\mathbf{P}$, through several transient intermediates $\mathbf{I}$, constitutes a full CRM as shown in eqn(1).
\begin{gather}
    \mathbf{R}\xrightarrow{step   S_{1}}I_{1}\xrightarrow{step S_{2}}I_{2},{\dotsm},{I_{n}}\xrightarrow{step S_{n}}\mathbf{P}
\end{gather}
where $S_{1:n}$ represents individual elementary steps. 

\noindent\textbf{Reactive atom.} Reactive atoms are the active atoms involved in an elementary step that undergoes a large change in their immediate bonding/valency environment as a result of the reaction.

\noindent\textbf{Reaction templates.} Reaction templates are defined as predetermined sets of chemical transformation rules with specific constraints, such as the presence of a particular substructure. \cite{landrum2013rdkit} In Figure~\ref{fig1:Method}a, an illustration of a template is provided using oxidative addition as a representative elementary step in a reaction. The specified constraints for this template are that any eligible reactant should possess a substructure featuring an aryl C-X bond (where X = Cl, Br, I) and that a catalytically active palladium (Pd) metal center be present.

\subsection{CRM prediction via ReactAIvate}

We develop an interpretable GNN model, ReactAIvate, to predict the elementary steps, which is further used to devise CRM. An extensive dataset comprising of seven such mechanistic steps for transition metal-catalyzed reactions is curated (see Section~\ref{subsec:dataset} for further details). ReactAIvate is build upon two elementary tasks:

\noindent\textbf{Reaction step classification (RSC).} The primary task of ReactAIvate is to predict the correct elementary step among the seven identified elementary steps. These mechanistic steps include `oxidative addition', `boron transmetallation', `acid-base deprotonation', `boronate formation', `substrate coordination', `transmetallation', and `reductive elimination'.\cite{kurosawa2003fundamentals} Once the correct mechanistic step is classified, an off-the-shelf template based reaction rules are used to predict the product information accurately. The predicted products form the reactants for the next step, and the subsequent mechanistic step is identified again. The process is repeated until the catalyst is regenerated. Our advantage in CRM prediction stems from our focus on mechanistic step classification, providing a distinct edge over traditional sequence-based modeling approaches. More specifically, given the set of reactants $\mathbf{R},$ or intermediates $I_{1}$, $I_{2}$, ..,$I_{n}$, ReactAIvate predicts labels denoted as $\{S_{i}\}_{i=1}^7$, where each $S_{i}$ corresponds to one of the seven mechanistic steps (Figure~\ref{fig1:Method}c).

However, while the dataset consists of the seven elementary steps, it doesn't encompass the entire range of mechanistic rules that a set of reactants may undergo in a chemical reaction. To ensure ReactAIvate doesn't erroneously predict reactions following different chemical transformation rules or force chemically non-reactive combinations into the seven elementary steps, we introduce an eighth class, denoted as $S_8$. Our framework is trained to classify any out-of-distribution (OOD) samples into this eighth class, bolstering confidence in ReactAIvate's predictions.


\noindent\textbf{Reactive atom identification (RAI).} Reactive step classification alone falls short in revealing the fundamental mechanisms and rationales underlying a CRM. Accurate identification of reaction centers within the reactant molecules is important for forward synthesis. This approach is particularly crucial when dealing with OOD samples, where the identification of reactive atoms within a given reaction class holds significant value. These insights serve as a guide for experts in recognizing feasible elementary reactions. In this work, the labels for reactive atom classification are derived from reaction templates, with `1' indicating a reactive atom and `0' indicating a non-reactive atom.

\subsubsection{ReactAIvate Workflow}
\label{subsec:workflow}
We now proceed to outline the operational methodology of our framework, ReactAIvate.

\noindent \textbf{1. Molecular representaion via graphs} A molecule can be portrayed as a graph, where atoms and bonds constitute nodes and edges, respectively \cite{meng2023unified}. Each unique atom is represented by the following set of features, encompassing nine types such as atom symbol, formal charge, hybridization, aromaticity, etc. These collectively result in a total of 39 atom features (see supporting information for further details). Similarly, edge representation includes feature vectors corresponding to different bond types (single, double, triple, etc.).

In our graph representation, $G = (V, E)$, where $V$ is the set of atoms and $E$ is the set of edges, each atom $\nu$ is associated with a feature vector $X_{\nu}$ in $\mathbb{R}^D$, with $D = 39$ representing the number of features for each atom. The tasks at hand involve: (1) classifying elementary steps into predefined classes. Given an input graph $G$ consisting of reactant molecules, the goal is to learn the representation vector $h^*_{G}$ and a linear function $g_1$ such that the predicted step $\hat{y}_G \coloneqq g_1(h^*_{G})$ aligns with the true step $y_G \in \{S_i\}{i=1}^8$; (2) identifying reactive atoms, where each atom $\nu$ in $V$ has a label $y_{\nu}$ ($y_{\nu} \in [0,1]$). The objective is to learn the representation vector $h^*_{\nu}$ and a linear function $g_2$ for all $\nu$ such that the predicted reactivity $\hat{y}_\nu \coloneqq g_2(h^*_{\nu})$ aligns with the true binary label $y_\nu$.  

\noindent \textbf{2. Graph attention network for RSC \& RAI} After encoding reactant molecules as graphs, we introduce an attention mechanism to these graphs by generating a context vector for a target atom $(\nu)$ through attention on its neighboring atoms ${u}$, where $u\in\mathcal{N}(\nu)$. The initial step in computing the context vector involves determining attention weights $\alpha_{u\nu}$ between the state vectors $h_{\nu}$ and $h_{u}$ of the two atoms~\cite{velivckovic2017graph,xiong2019pushing}. Subsequently, a context operation follows, where a linear transformation is applied to $h_{u}$, the state vectors of neighboring atoms. This is succeeded by a weighted sum and a non-linear activation function, resulting in $C_{\nu}$, the context vector for the target atom $\nu$. The calculation of normalized attention coefficients can be formulated as follows: 
\begin{equation*}
    \alpha_{u\nu} = \frac{\exp\left(\texttt{LeakyReLU}\left(\textbf{W}[{h}_{\nu},{h}_u]\right)\right)}{\sum_{u \in \mathcal{N}(\nu)} \exp\left(\texttt{LeakyReLU}\left(\textbf{W}[{h}_{\nu},{h}_u]\right)\right)},
\end{equation*}
where $\alpha_{u\nu}$ signifies the importance (weight) of neighbor atom $u$ to target atom $\nu$, and \textbf{W} is a trainable weight matrix, with
\begin{equation*}
    C_{\nu} = \texttt{ELU} \left({\sum_{u \in \mathcal{N}(\nu)} \alpha_{u\nu}.\textbf{W}.h_u}\right).
\end{equation*}

Modern GNNs adopt neighborhood aggregation strategies, updating the features of the target atom iteratively by incorporating the features of its neighboring atoms. The atom's representation encapsulates the structural information within the $k$-hop network around it after $k$ iterations of this aggregation process. This strategy can be formulated as:       

\noindent\textbf{Aggregation phase}
\begin{equation*}
    C^{k-1}_{\nu} = {\sum_{u \in \mathcal{N}(\nu)}A^{k-1}(h^{k-1}_u,h^{k-1}_\nu)},
\end{equation*}

\noindent\textbf{Update phase}
\begin{equation*}
    h^{k}_{\nu} = {\texttt{GRU} ^{k-1}(C^{k-1}_{\nu},h^{k-1}_\nu)},
\end{equation*}
where, $h^{k}_{\nu}$ represents the feature vector of atom $\nu$ after the $k$th layer, with $h^{0}{\nu} = X_{\nu}$ (see Figure~\ref{fig1:Method}c). In the aggregation phase, the graph attention mechanism, $A^{k-1}$, provides the most relevant information to the target atom from its neighborhoods in the form of the context vector $C^{k-1}_{\nu}$. In the subsequent step, the update function $\texttt{GRU}^{k-1}$ (gated recurrent unit) takes the attention context and the previous state vector of the target atom as input to update the feature vector from the previous state $h^{k-1}_{\nu}$ to the current state $h^{k}_{\nu}$. 

To generate a graph-level embedding, a virtual \emph{supernode} is introduced, connecting with all the atoms in the molecular graph. The feature vector for this supernode is obtained using sum pooling, expressed as $h^{0}_{s} = \sum_{\nu}h^{k}_{\nu}$. Through a similar neighborhood aggregation method, a high-level graph embedding $h^{t}_{s}$ is learned iteratively over $t$ iterations. At this point, the optimally learned vector $h^{t}_{s}$ can be considered equivalent to $h^*_{G}$ (see Figure~\ref{fig1:Method}c), which is subsequently utilized for the elementary step prediction task. By feeding the graph-level embedding $h^{*}_{G}$ into a feed-forward neural network (FNN), predicted values are obtained for the RSC task as:
\begin{equation*}
    \hat{y}_G = \texttt{FNN}(h^{*}_{G}).
\end{equation*}

ReactAIvate employs cross-entropy loss between the predicted and true labels for the RSC task:
\begin{equation*}
    \mathcal{L}^G_{class} \coloneqq \texttt{CrossEntropyLoss}(\hat{y}_{G},{y}_{G}).
\end{equation*}

In our RAI task, the objective is to classify each atom in a molecule as reactive or non-reactive, enabling the use of standard classification loss functions. ReactAIvate employs binary cross-entropy (BCE) loss for two-way classification, with a slight modification. Given that the majority of atoms in a sample are non-reactive, we modify the BCE loss by introducing a weight that strongly penalizes incorrect predictions for atoms that are genuinely reactive. This adjustment helps prevent the model from becoming biased towards predicting all atoms as non-reactive (see supporting information).

Recall that for an atom $\nu$, the associated embedding vector before the sum pooling is denoted as $h^{k}_{\nu}$. This embedding vector is considered the optimal atom representation $h^*_{\nu}$. In the node-level classification task, the updated node feature vector is passed through the FNN to predict whether the atom is reactive or non-reactive, expressed as:
\begin{equation*}
	\hat{y}_{\nu} = \texttt{FNN}(h^*_{\nu}).
\end{equation*}
For RAI, we consider the weighted BCE loss as discussed above:
\begin{equation*}
    \mathcal{L}^G_{RC} \coloneqq \sum_{\nu}\texttt{WeightedBCELoss}(\hat{y}_{\nu},{y}_{\nu})
\end{equation*}

The overall loss for ReactAIvate is made up of the individual losses for RSC and RAI: 
\begin{equation*}
    \mathcal{L} = \mathcal{L}^G_{class} + \mathcal{L}^G_{RC}
\end{equation*}

\section{Experiments}

\subsection{Dataset details}
\label{subsec:dataset}

In this study, we consider a diverse and representative set of transition metal-catalyzed reactions. These include Suzuki-Miyaura coupling (SMC) \cite{martin2008palladium,beletskaya2019suzuki}, Buchwald-Hartwig amination (BHA) \cite{ruiz2016applications,dorel2019buchwald}, and Kumada coupling (KC) \cite{ackermann2006air}. The inclusion of these reactions is motivated by their significance in drugs, agrochemicals, and pharmaceutical synthesis. To the best of our knowledge, there is currently no existing database that includes credible mechanisms for these reactions, compelling the creation of CRM datasets. For this, seven distinct elementary mechanistic steps are recognized that can account for all three reaction mechanisms: `oxidative addition', `boron transmetallation', `acid-base deprotonation', `boronate formation', `substrate coordination', `transmetallation', and `reductive elimination' (see supporting information).

We have created essential reaction templates for each elementary mechanistic step within the three considered reaction datasets in this study. The template is structured to yield the product based on the given reactants (refer to Figure~\ref{fig1:Method}a). A reaction comprises a set of substrates, catalyst and reagents. For example, in the case of SMC, key substrates include an aryl halide and a boronic acid, with a metal-phosphine complex serving as the catalyst and a base facilitating the reaction. These reaction partners are curated from primary literature \cite{wolfe1997improved,huang2003expanding}and the PubChem database \cite{kim2021valid}. By inputting this set of molecules into the reaction template aligned with the specific reaction mechanism type, all elementary steps and the complete CRM can be derived. Our dataset encompasses a total of 100,000 elementary mechanistic steps (see supporting information for more details).

\subsection{Training details} 

The dataset is partitioned into training, validation, and test samples with a distribution ratio of 70:10:20. ReactAIvate is constructed using PyTorch \cite{paszke2019pytorch} with the Adam optimizer \cite{kingma2014adam} and a batch size of 256. Throughout this study, we maintain consistent hyperparameter values: $k$ (number of attentive message passing layers for atom embedding) is set to 2, $t$ (number of attentive message passing layers for molecule embedding) is set to 1, with an $L2$ weight decay of 0.000001, a learning rate of 0.001, and a dropout rate of 0.1. The number of atom features and graph feature size are specified as 39 and 200, respectively. 

\subsection{Baseline}

First, we employed Seq2Seq approaches, such as  T5Chem~\cite{lu2022unified} and Transformer~\cite{rush2018annotated} as proxies for baselines, which are considered state-of-the-art (SOTA) for individual single step reaction prediction. T5Chem is a multi-tasking model designed for various reaction prediction tasks, leveraging the Text-To-Text Transfer Transformer (T5), an encoder-decoder model from the transformer family. The Transformer baseline is built on the original encoder-decoder framework~\cite{vaswani2017attention}. 

In these frameworks, the problem is formulated as a generation task, where the model is trained to predict the full CRM given the reactants and other entities. To ensure a fair comparison, we introduce two test datasets, each comprising 1000 samples. The first is an in-distribution (ID) test dataset, where the individual reacting partners of each sample belong to the same set as the training datasets, although the samples themselves are not part of the training set. The second is an out-of-distribution (OOD) dataset, where the reaction components of each sample differ structurally from those used in training the model.

\begin{table}
    \centering
    \small
    \begin{tabular}{p{1.8cm}ccc}
        \toprule
        \multirow{2}{*}{Model} & \multicolumn{2}{c}{Accuracy (\%)} & {Train} \\
         & ID test set & OOD test set & time (m) \\
        \midrule
        \textit{\bf Seq2Seq}  &    &   &   \\
        T5Chem  & 95.60 $\pm$ 0.01 & 0.07 $\pm$ 0.01 & 180 \\
        T5Chem(FI)  & 98.40 $\pm$ 0.01 & 0.11 $\pm$ 0.02 & 45 \\
        Transformer & 11.10 $\pm$ 0.03 & 0.01 $\pm$ 0.00 & 150 \\ 
        Transformer(FI) & 91.80 $\pm$ 0.05 & 0.07 $\pm$ 0.00 & 60 \\ 
        \midrule
        \textit{\bf Featurization}  &    &   &   \\
        Morgan(r=2) & 100.00 $\pm$ 0.00 & 31.50 $\pm$ 3.25 & 20 \\
        MFF & 100.00 $\pm$ 0.00 & 48.76 $\pm$ 3.60 & 30 \\
        PCD & 100.00 $\pm$ 0.00 & 48.57 $\pm$ 2.67 & 10 \\
        \textit{Graph}  &    &   &   \\
        MPNN & 100.00 $\pm$ 0.00 & 52.80 $\pm$ 9.50 & 10 \\
        ReactAIvate &  $\mathbf{100.00\pm 0.00}$ &  $\mathbf{95.70\pm 0.34}$ & $\mathbf{15}$ \\
        \bottomrule
    \end{tabular}
    \caption{Performance of different models for CRM prediction on ID and OOD test molecules. Training time is reported in minutes (m). \textbf{Bold} indicates best performance.}
    \label{table:CRMPred}
\end{table}


Next, we developed several DNN models as baselines that are not based on sequence data such as in seq2seq models. These models use different feature representations, including fingerprint (Morgan and multiple fingerprint feature (MFF)) ~\cite{sandfort2020structure}, and rdkit-based physicochemical descriptors (PCD)) ~\cite{riniker2013open}. Their design targets the same objective as ReactAIvate. Additionally, we considered a Message-Passing Neural Network (MPNN) model as another benchmark ~\cite{gilmer2017neural}. It operates by passing messages between nodes (atoms) in a graph (molecule) to learn the features and interactions. More details about baseline models are provided in the supporting information.

\begin{figure*}[!ht]
	\begin{center}
		\begin{tabular}{c}
			\includegraphics[width=2.0\columnwidth]{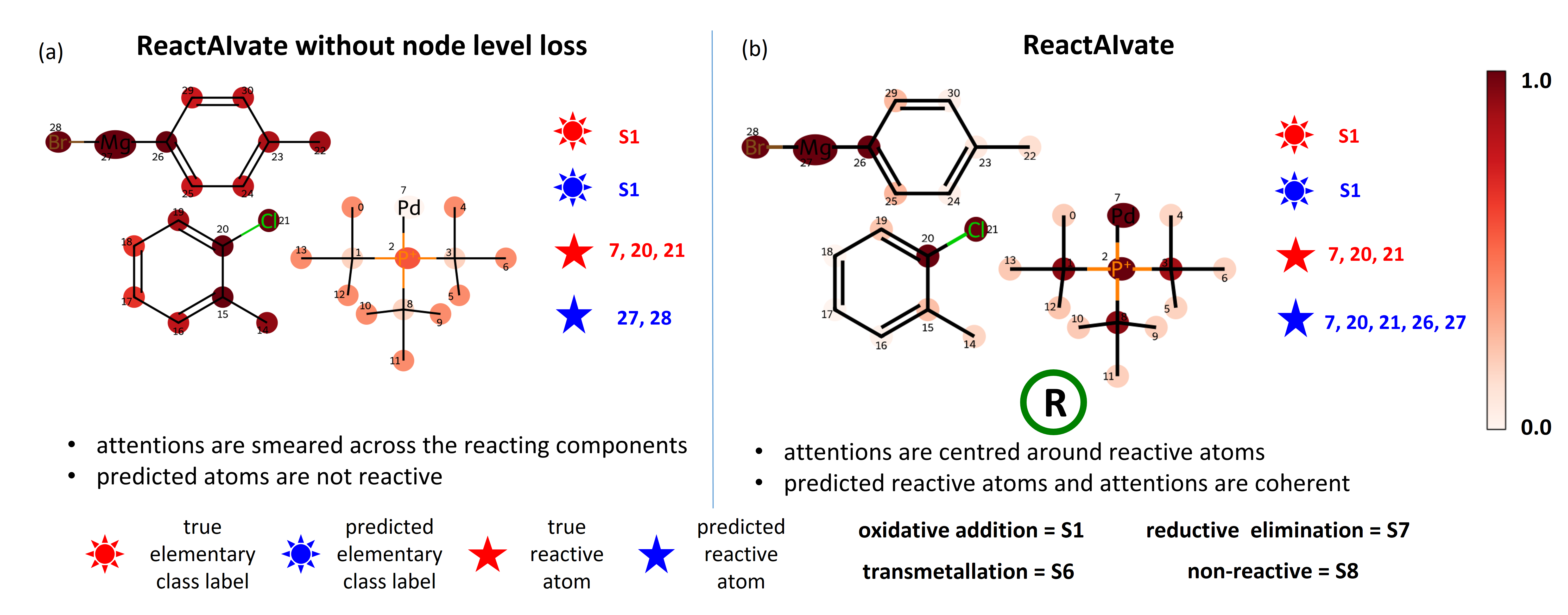}
		\end{tabular}
		\caption{Effect of the inclusion of node-level loss in ReactAIvate demonstrated through attention visualization. The rightmost bar represents min-max rescaled attention values.}
		\label{fig:NodeLoss}
	\end{center}
\end{figure*}

\begin{figure*}[!ht]
	\begin{center}
		\begin{tabular}{c}
			\includegraphics[width=2.0\columnwidth]{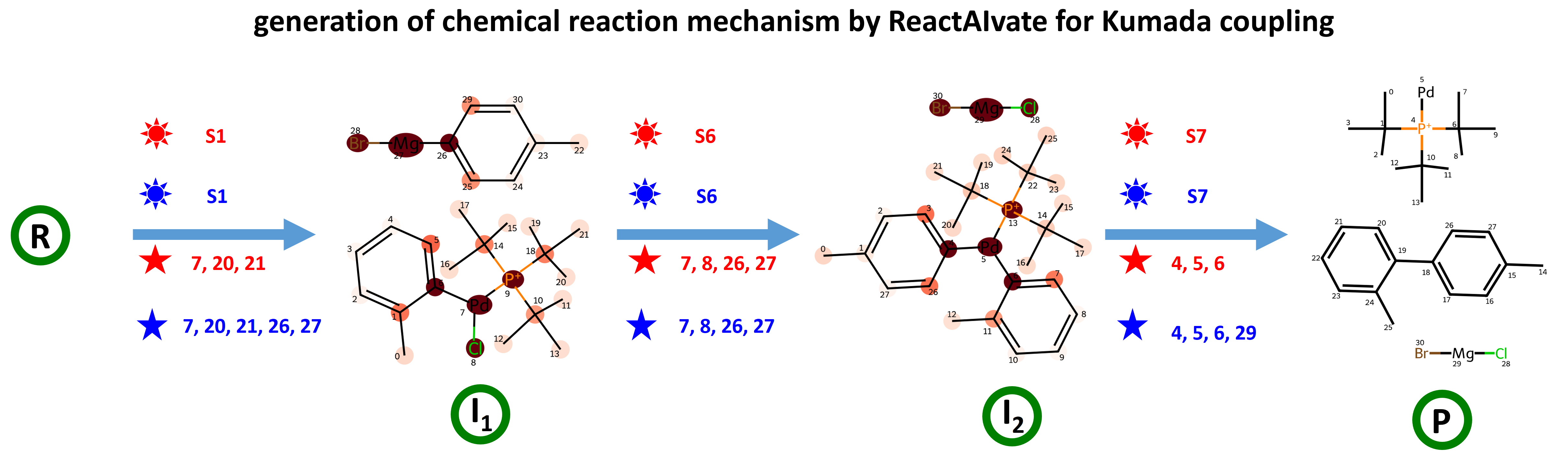}
		\end{tabular}
		\caption{An illustration of the sequential generation of the full CRM for the Kumada coupling reaction}
		\label{fig:Illustration}
	\end{center}
\end{figure*}

\begin{figure*}[!ht]
	\begin{center}
		\begin{tabular}{c}
			\includegraphics[width=2.0\columnwidth]{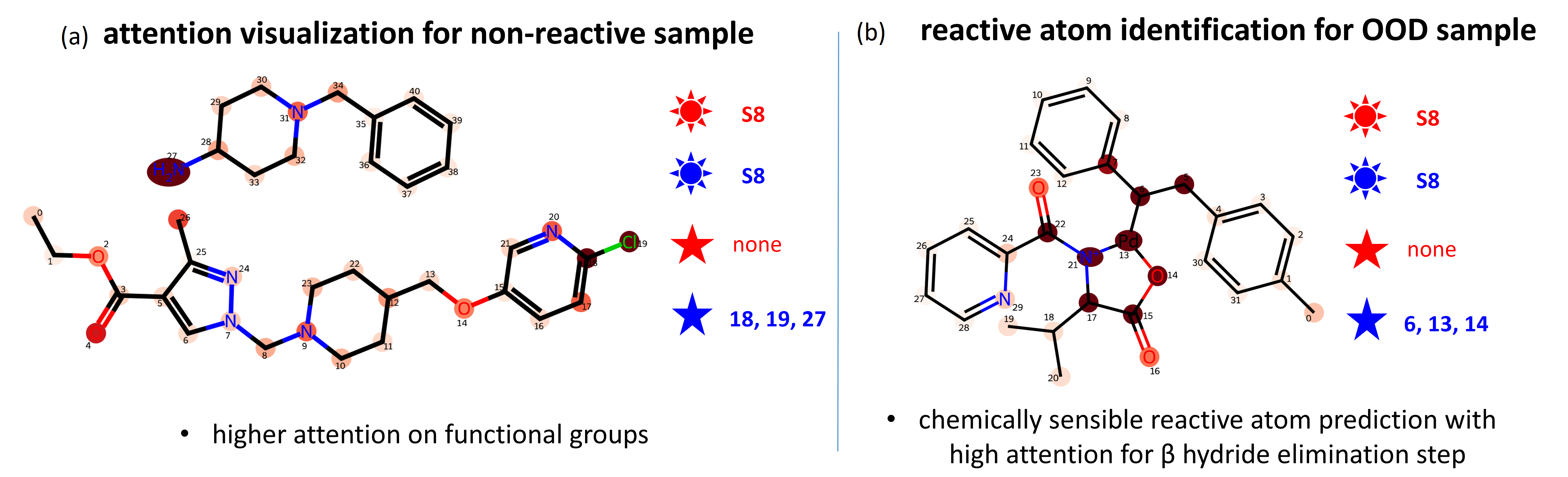}
		\end{tabular}
		\caption{Attention visualization for a sample in (a) non-reactive, (b) reactive out-of-distribution (OOD) set.}
		\label{fig:OOD}
	\end{center}
\end{figure*}

\subsection{Results}
Table~\ref{table:CRMPred} illustrates the performance comparison, specifically focusing on accuracy, i.e., the percentage of samples where the generated CRM precisely matches the true CRM. For the ID dataset, ReactAIvate exhibits better accuracy compared to T5Chem and Transformer. Intriguingly, in the OOD dataset, ReactAIvate offers superior accuracy in CRM generation, reaching upto 96\%. In contrast, both the baselines fail to predict the correct CRM, potentially due to incorrect predictions of one of the many tokens or atom imbalance, rendering the entire sequence invalid (see supporting information for more details). The observation is not specific to CRM prediction, but has also been reported in recent findings, where several SOTA template-free models unexpectedly falter when faced with OOD in retrosynthesis prediction~\cite{chen2024assessing}. In addition, unlike the above-mentioned baselines, where T5Chem and Transformer-based architectures are trained to predict the entire CRM, we have conducted additional experiments where the baselines are trained to predict just the forward intermediates (FIs) at each step (exactly what these models were originally designed for), and then the predicted intermediates serve as the reactants for the next step. The process is repeated until the catalyst is regenerated (i.e., end of CRM). The results are reported above in Table~\ref{table:CRMPred}. Even for the FI variants of the baselines, their performances remain highly inadequate for OOD samples. Although T5Chem benefits from pre-training on the Pubchem database containing  \textbf{97M molecules}, its sequence-centric framework limits its effectiveness in CRM prediction, especially with OOD instances.

The poor OOD performance of seq2seq/transformer models is not due to poor hyperparameter tuning but rather the nature of character generation-based product prediction. For example, if a model is trained on molecules with Iodine and Chlorine but tested on a molecule where Iodine is replaced with Fluorine, it struggles because it was never trained to generate sequences involving Fluorine. Additionally, any incorrect symbol prediction in an intermediate step can corrupt the entire downstream task for seq2seq models. On the contrary, ReactAIvate predicts the most appropriate reaction rule (template) at each step, using these rules to accurately generate the right products at each step.

The fingerprints and PCD based DNN models exhibited enhanced performance as compared to sequence-based models on both ID and OOD samples. However, it is worth noting that these models yield lower accuracy for the OOD set as compared to ReactAIvate, this discrepancy may stem from the fact that fingerprints rely on predefined sets of molecular substructures, potentially limiting their ability to capture nuanced structural details. Whereas MPNN performed better than fingerprints and PCD featurization, they still fell short of ReactAIvate. This could be attributed to the advantages conferred by attention mechanisms and the incorporation of RAI task within ReactAIvate. This, in turn, implicitly captures the relevance between CRM and RAI, and also helps with the alignment of highly attentive atoms with likely reaction centers. Nonetheless, these models provide robust and adequate validation. These results strongly imply the necessity for a meticulous reconsideration of the modeling aspect for CRM generation.

\begin{table}
    \centering
    \small
    \begin{tabular}{lcc}
        \toprule
        \textbf{Task} & \textbf{Accuracy} (\%) & \textbf{F1-Score}\\
        \midrule
        ID-RSC & 100.00 $\pm$ 0.00 & 1.00 $\pm$ 0.00\\
        ID-RAI & 96.40 $\pm$ 0.30 & 0.87 $\pm$ 0.01 \\
        OOD-RSC & 98.60 $\pm$ 0.23 & 0.98 $\pm$ 0.01\\
        OOD-RAI & 94.86 $\pm$ 0.40 & 0.85 $\pm$ 0.02\\
        \bottomrule
    \end{tabular}
    \caption{Performance of ReactAIvate for RSC and RAI tasks on ID and OOD test molecules.}
    \label{table:ReactAIvate}
\end{table}

We delve deeper into ReactAIvate's capability for accurately predicting RSC and RAI concurrently. The classification accuracies and F1-scores for ID and OOD reactants are presented in Table~\ref{table:ReactAIvate}. Beyond accurately classifying the correct elementary step, ReactAIvate demonstrates proficiency in distinguishing between reactive and non-reactive atoms. Notably, even in the OOD dataset, ReactAIvate maintains high efficacy for both RSC and RAI tasks. This indicates that the model has a broad applicability domain and holds potential utility for domain experts.

\section{Discussion and Analysis}
Having demonstrated the exceptional performance of our framework ReactAIvate for both the RSC and RAI tasks, we proceed to delve into the underlying reasons for its effectiveness through a comprehensive ablation study. Additionally, we aim to elucidate domain-specific advantages that contribute to its superior performance.

\noindent\textbf{Significance of incorporating node-level RAI}: We begin by emphasizing the implication of integrating the node-level loss in the reactive atom prediction task along with the graph-level loss for elementary step prediction. In Figure~\ref{fig:NodeLoss}, we compare the attention visualization between ReactAIvate without the node-level loss and ReactAIvate with both losses, using an oxidative addition step from the Kumada coupling dataset. In the former model, attention weights are dispersed across the reactants and catalyst, providing limited utility. Notably, the crucial reactive metal atom Pd:7, responsible for catalyzing this step, fails to draw any attention. Thus, attention visualization without the node-level loss contributes little to understanding chemical reactivity. In the latter model with the two-level loss, attention weights are more concentrated around the reactive region of the molecules. For instance, the Pd metal center and the aryl chlorine bond garner higher attention, (Pd:7, C:20, Cl:21), representing the true reactive atoms in this elementary step. This underscores the importance of including the node-level loss, aligning the model's attention mechanism more closely with how a chemist would assign attention to reactive atoms.

\noindent \textbf{Illustration of a full CRM}: Moving forward, we assess ReactAIvate's capability in predicting a complete CRM for Kumada coupling, serving as a representative example (Figure~\ref{fig:Illustration}). In the initial step, the model accurately predicts oxidative addition ($S_1$) as the elementary step, along with the correct identification of reactive atoms. Notably, the model also predicts two additional atoms as active, which intriguingly turn out to be reactive in the subsequent step. In the second step, ReactAIvate once again correctly predicts the elementary class, transmetallation ($S_6$), and accurately identifies all true reactive atoms. Finally, the model predicts the reductive elimination ($S_7$) step as the concluding phase. Both the predicted active atoms and attention distributions align consistently with the expected mechanism. In summary, ReactAIvate demonstrates the ability to generate a complete CRM starting from only reactants and catalysts.

\noindent \textbf{Attention visualization for OOD-class}: To gain deeper insights into the model, we visualize attentions in a sample belonging to the OOD class, as shown in Figure~\ref{fig:OOD}a. In this instance, the pair of molecules is predicted to fall into the eighth (OOD) class, accurately reflecting that the amine and the aryl halide (N:27, C:18, Cl:19) would not react in the absence of a catalyst. Intriguingly, the model predicts chlorine in one molecule and the adjacent carbon in the other as reactive atoms. Additionally, attentions are primarily distributed around functional groups. The broad dispersion of attention underscores ReactAIvate's challenge in pinpointing the exact reaction mechanism. This outcome aligns with expectations, as the combination is chemically non-reactive, and it would be counterintuitive for the model to highlight specific reaction centers leading to its prediction in the OOD class.

In our final evaluation, we test ReactAIvate for the identification of potential reactive atoms in an OOD sample involving an entirely different reaction mechanism, as depicted in Figure~\ref{fig:OOD}b. Please note that the molecule is not expected to undergo any transformation (both predicted and true class are $S_8$, i.e., no reaction). However, if it is presented with a suitable reagent, this molecule is anticipated to undergo a $\beta$ hydride elimination step and the reactive centers get activated. Among the predicted reactive atoms, two (Pd:13 and  $\alpha$-C: 6) correspond to possible true reactive atoms. Since the molecule in this example doesn't undergo reaction, the true reactive centers are presented as an empty list. Moreover, attentions are dispersed around the reaction center and the reactive atoms. These highlights the model's versatility in predicting OOD samples and serves as a guide for comprehending the reactivity of metal-catalyzed reactions. Consequently, one can easily incorporate new elementary classes of interest to further broaden the model's applicability.

\section{Conclusion and Future Work}
In conclusion, we introduce ReactAIvate, a graph-attention-based Graph Neural Network (GNN) model designed for interpretable Chemical Reaction Mechanism (CRM) generation. Our model is trained on a novel dataset comprising seven distinct elementary mechanistic steps, covering the complete CRM for three different transition-metal-catalyzed processes. ReactAIvate excels in accurately classifying elementary steps and recognizing reactive atoms, demonstrating its capability to construct full CRMs. Notably, the model exhibits a prudent handling of non-reactive cases, showcasing its reliability in predictions. ReactAIvate outperforms Seq2Seq baseline models, emphasizing the limitations of the latter in CRM identification due to minor errors. The robust OOD classification performance underscores ReactAIvate's potential for exploring additional mechanisms with the availability of more data. As part of future work, we plan create a user-friendly interface for predicting entire CRMs based on user-inputted SMILES of reactants. 

\section{Data Availability}

Data and codes related to this work are publicly available through our Github repository at \url{https://github.com/alhqlearn/ReactAIvate}.

\section{Acknowledgements}

We gratefully acknowledge the generous computing time provided by the SpaceTime supercomputing facility at IIT Bombay. M.D. expresses gratitude for the Prime Minister’s Research Fellowship.


\bibliography{mybibfile}

\end{document}



\begin{frontmatter}


\paperid{1850} 


\title{ReactAIvate: A Deep Learning Approach to Predicting Reaction Mechanisms and Unmasking Reactivity Hotspots}


\author[A]{\fnms{Ajnabiul}~\snm{Hoque}\orcid{0000-0001-9807-3061}\footnote{Equal contribution.}}
\author[A]{\fnms{Manajit}~\snm{Das}\orcid{0000-0001-7709-8809}\footnote{Equal contribution.}}
\author[B,C]{\fnms{Mayank}~\snm{Baranwal}\thanks{Corresponding Author. Email: baranwal.mayank@tcs.com}\orcid{0000-0001-9354-2826}}
\author[A,D]{\fnms{Raghavan}~\snm{B.	Sunoj}\thanks{Corresponding Author. Email: sunoj@chem.iitb.ac.in}\orcid{0000-0002-6484-2878}} 

\address[A]{Department of Chemistry, Indian Institute of Technology Bombay, India}
\address[B]{Department of Systems \& Control Engineering, Indian Institute of Technology, India}
\address[C]{Tata Consultancy Services Research, Mumbai, India}
\address[D]{Centre for Machine Intelligence and Data Science, Indian Institute of Technology Bombay, India}



\end{frontmatter}

\tableofcontents

\pagebreak

\section{Code details}
We provide our code, datasets, and pre-trained models via the following Github link-

https://github.com/alhqlearn/ReactAIvate 

\section{Elementary steps}

\begin{figure*}[!ht]
    \centering
    \includegraphics[width=0.8\linewidth]{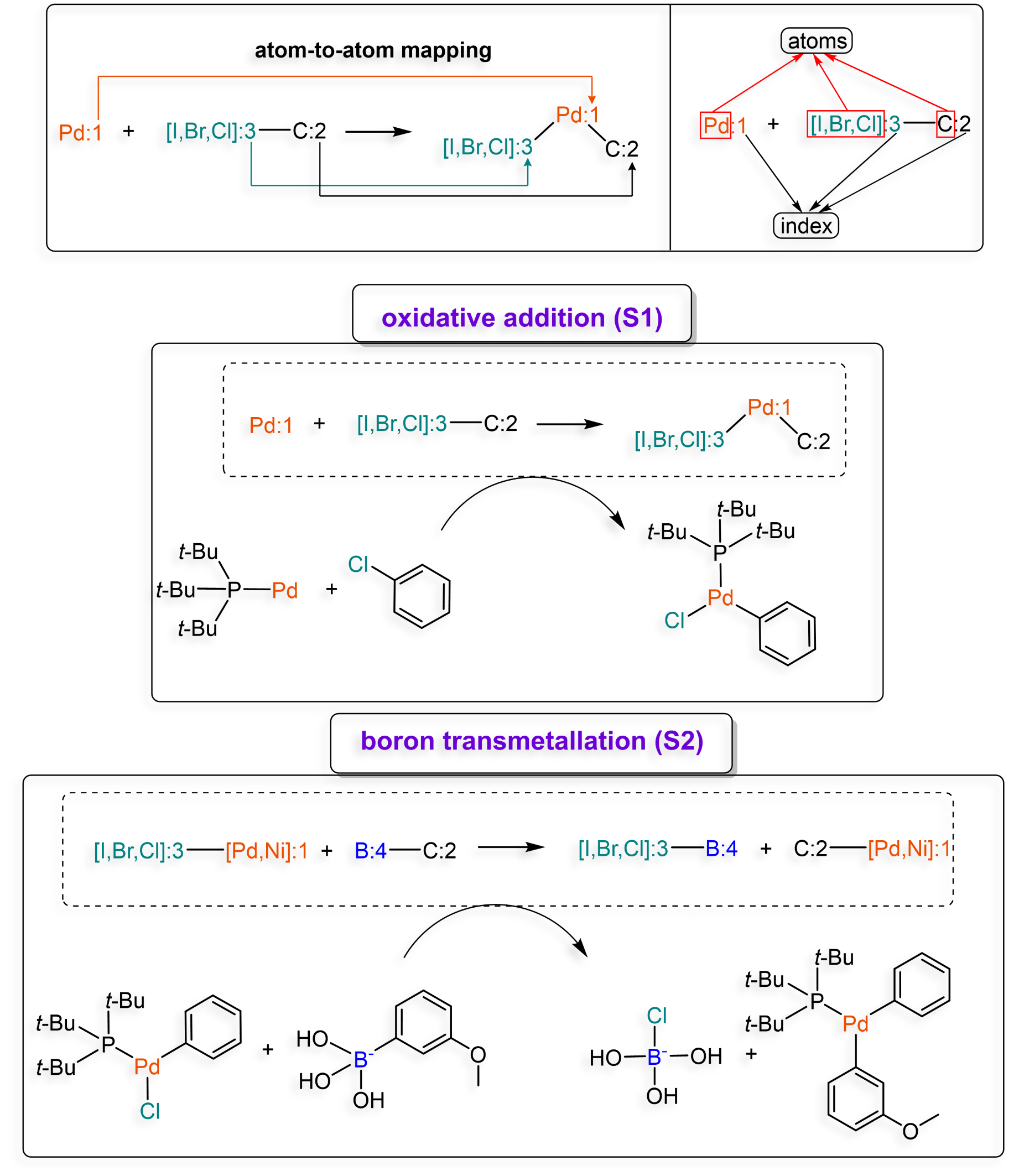}
    \caption{Atom-to-atom mapping (left) and atom specification with index (right) in a reaction template is shown in the first box. Representative examples of `oxidative addition' (S1) and `boron transmetallation' (S2) templates.}
    \label{fig:fig1}
\end{figure*}

\begin{figure*}[!ht]
	\begin{center}
		\begin{tabular}{c}
			\includegraphics[width=0.8\columnwidth]
                {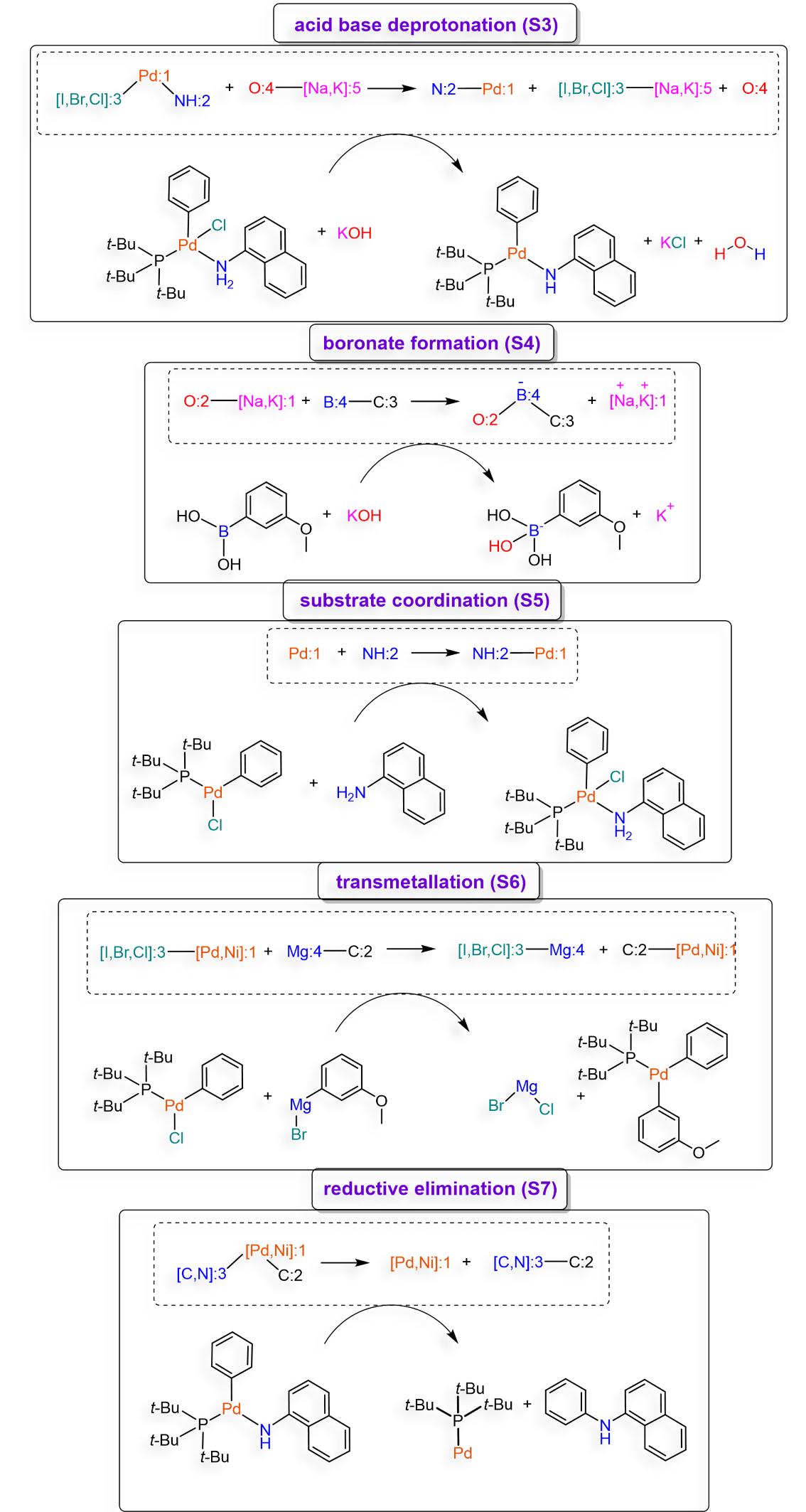}
		\end{tabular}
		\caption{Representative examples of different reaction templates `acid-base deprotonation' (S3), `boronate formation' (S4), `substrate coordination' (S5), `transmetallation' (S6), and `reductive elimination' (S7)}
		\label{fig:fig2}
	\end{center}
\end{figure*}

In this study, we created a novel dataset of important chemical reaction mechanisms (CRMs) featuring three distinct reaction classes, namely, Suzuki-Miyaura coupling (SMC) \cite{martin2008palladium,beletskaya2019suzuki}, Buchwald-Hartwig amination (BHA) \cite{ruiz2016applications,dorel2019buchwald}, and Kumada coupling (KC) \cite{ackermann2006air}. The dataset consists of seven unique elementary mechanistic steps, carefully chosen to capture the underlying mechanism of all three reactions. The corresponding reaction templates for these elementary steps, along with illustrative examples, are provided in Figures~\ref{fig:fig1},~\ref{fig:fig2}.

The reaction templates are based on specifying the bond changes from reactants to products along with the corresponding atom-to-atom mapping. This is illustrated using an oxidative addition step in the first box Figure \ref{fig:fig1}. The carbon (C:2), halogen ([I,Br,Cl:3]) single bond is broken, and two bonds (i) Pd:1 and [I,Br,Cl]:3 and (ii) Pd:1 and C:2 are formed. 

Similarly, in the boron transmetallation step two bonds are broken, namely, (i) the bond between boron and carbon i.e., B:4-C:2 and (ii) metal and halogen i.e., [Pd,Ni]:1-[I,Br,Cl]:3. Simultaneously, two new bonds are formed between (i) carbon and metal i.e., C:2-[Pd,Ni]:1 and (ii) boron and halogen i.e., B:4-[I,Br,Cl]:3. All the other templates can be described similarly.

\section{Chemical reaction example with mechanism}

\subsection{Suzuki-Miyaura coupling (SMC)}

\begin{figure*}[!ht]
    \centering
    \begin{tabular}{c}
        \includegraphics[width=0.8\columnwidth]{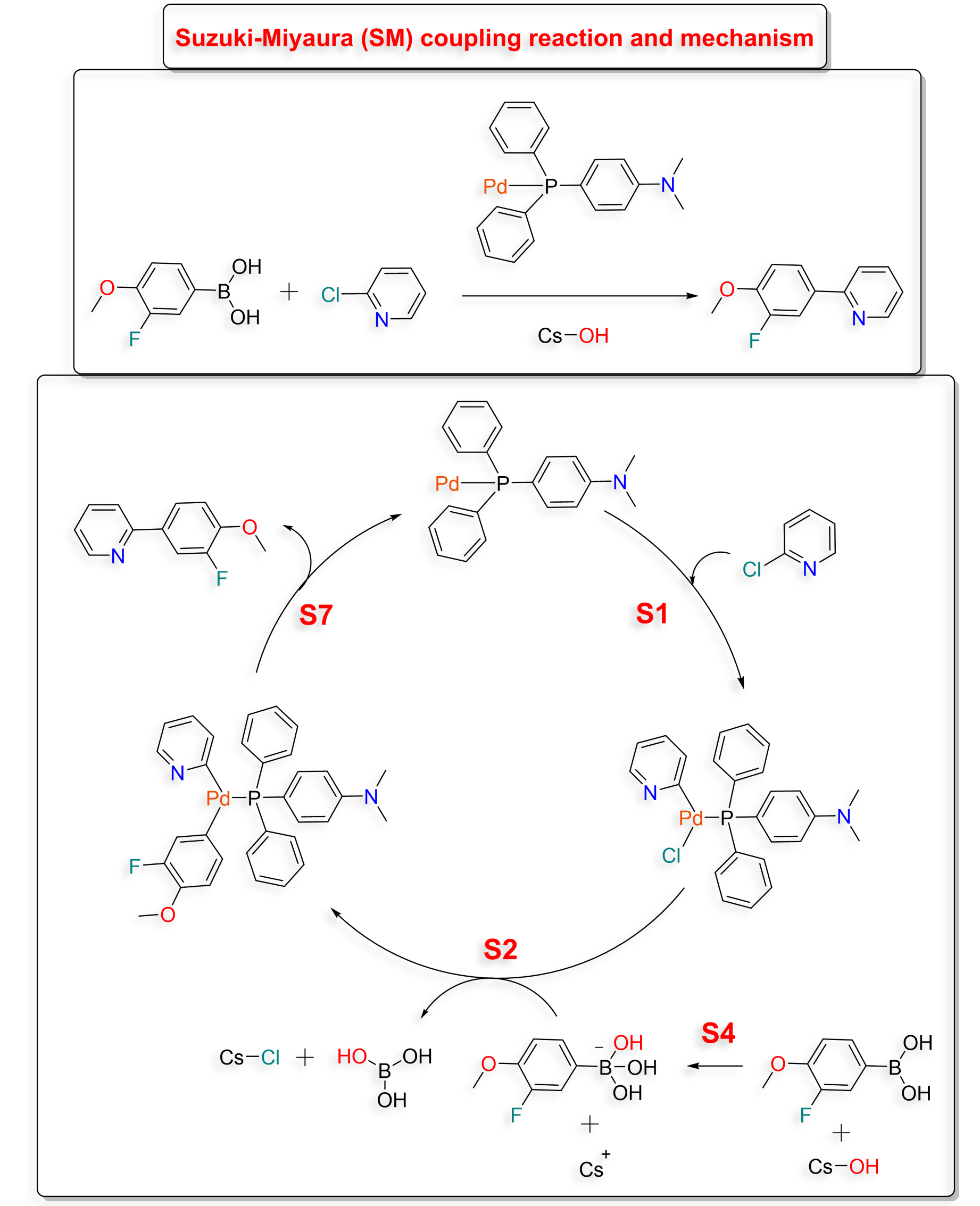}
    \end{tabular}
    \caption{An example of SMC reaction and the corresponding catalytic mechanism. The elementary steps are shown in red bold font. The reaction involves coupling of an aromatic boronic acid with aryl halide using palladium-phosphine based catalyst and a base}
    \label{fig:fig3}
    
\end{figure*}

The Suzuki–Miyaura coupling (SMC) is one of the most often employed carbon–carbon bond-forming reactions in the pharmaceutical industry \cite{miyaura2002cross,miyaura1995palladium}. 
This reaction class involves cross-coupling between organohalides and organoboron compounds in the presence of a Pd-catalyst and a suitable base. The mechanism of this reaction can be described by a sequence of four different elementary mechanistic steps. These are, i) oxidative addition, ii) boronate ester formation, iii) transmetalation, and iv) reductive elimination (see Figure \ref{fig:fig3}).

\subsection{Buchwald-Hartwig amination (BHA)}

\begin{figure*}[!ht]
	\begin{center}
		\begin{tabular}{c}
			\includegraphics[width=0.8\columnwidth]{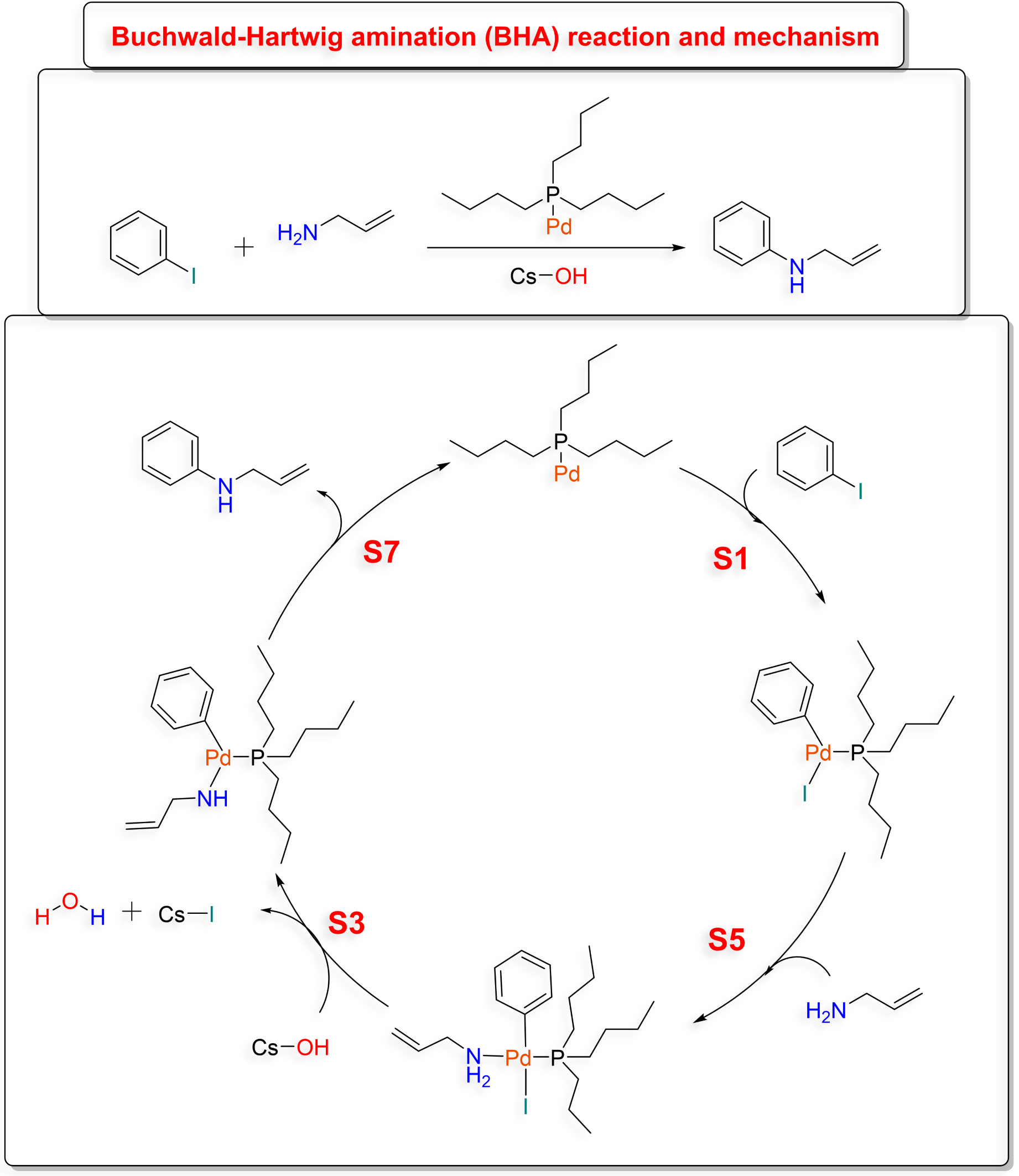}
		\end{tabular}
		\caption{An example of BHA reaction and the corresponding catalytic cycle/mechanism.  The elementary steps are denoted using red bold font. The reaction shown is a coupling between iodo benzene and allyl amine catalyzed by a palladium-phosphine catalyst in the presence of a base }
		\label{fig:fig4}
	\end{center}
\end{figure*}

Over the last two decades, the Buchwald-Hartwig (BH) amination or the palladium-catalyzed amination of aryl halides and pseudohalides has emerged as a valuable tool in organic synthesis, facilitating the creation of $C(sp^2)-N$ bonds \cite{forero201925th}. The underlying mechanism of this reaction involves four distinct elementary steps such as i) oxidative addition, ii) amine coordination, iii) acid-base deprotonation, and iv) reductive elimination (see Figure \ref{fig:fig4}).

\subsection{Kumada coupling (KC)}

\begin{figure*}[!ht]
	\begin{center}
		\begin{tabular}{c}
			\includegraphics[width=0.8\columnwidth]{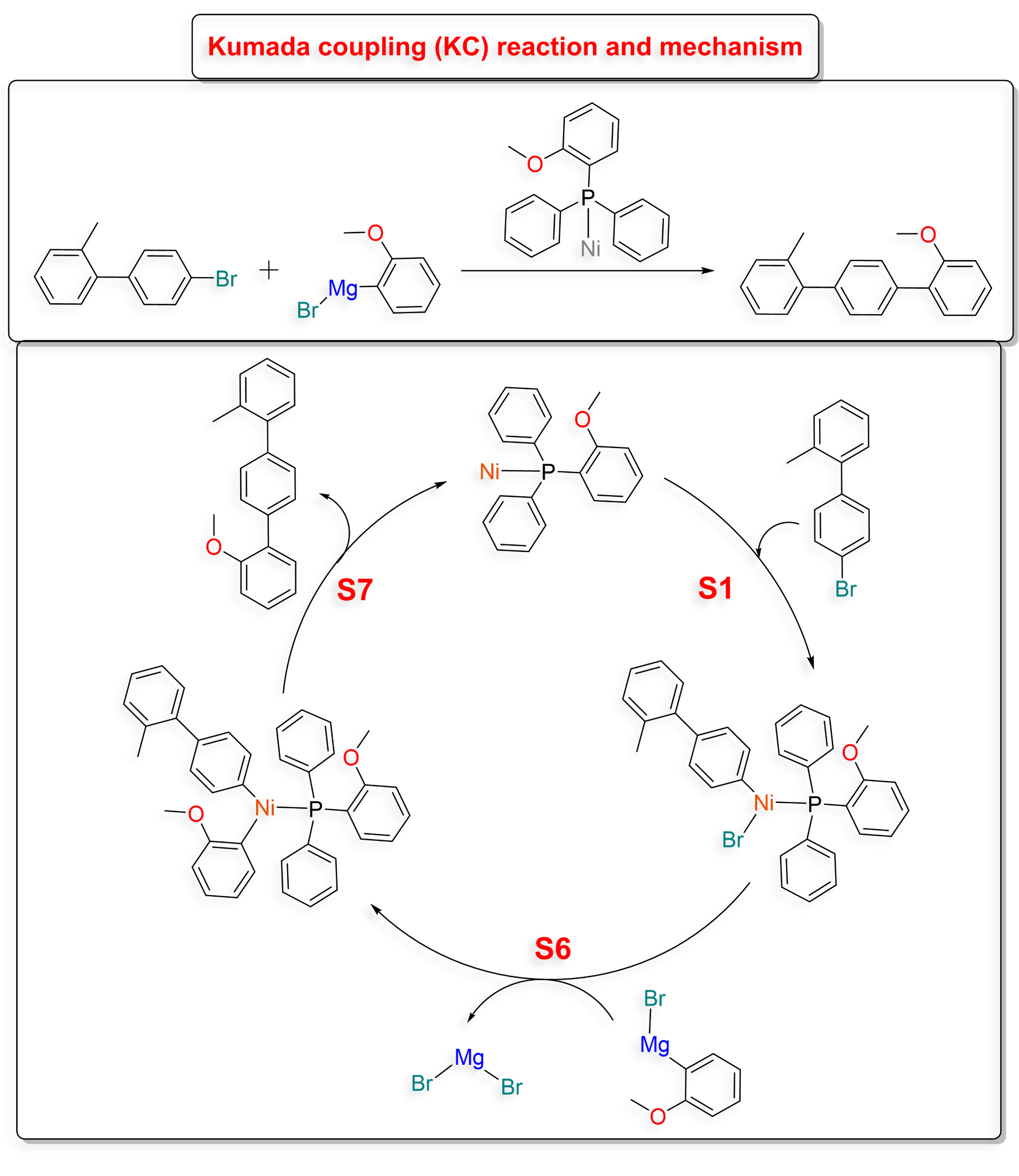}
		\end{tabular}
		\caption{An example of KC reaction and the corresponding mechanism. The elementary steps are denoted using red bold font. The reaction is shown for coupling of biaryl bromide with aryl magnesium bromide using nickel-phosphine based catalyst}
		\label{fig:fig5}
	\end{center}
\end{figure*}

The Kumada cross-coupling is a reaction between an organohalide and an organomagnesium compound (commonly known as a Grignard reagent). This reaction, catalyzed by a palladium or nickel catalyst, results in the formation of a $C-C$ coupled product. The catalytic cycle for this reaction follows a sequence of three elementary steps, i) oxidative addition, ii) transmetallation, and iii) reductive elimination (see Figure \ref{fig:fig5}).

\section{Atom and bond features}

\begin{table*}[!ht]
    \centering
    \small
    \begin{tabular}{lrll}
        \toprule
        atom feature & size & description & feature type \\
        \midrule
        atom symbol & 16 & B, C, N, O, F, Si, P, S, Cl, As, Se, Br, Te, I, At, metal & one-hot \\
        degree & 6 & number of covalent bonds [0, 1, 2, 3, 4, 5] & one-hot \\
        formal charge & 1 & electrical charge & integer \\
        radical electrons & 1 & number of radical electrons & integer \\
        hybridization & 6 & [$sp, sp^2, sp^3, sp^3d, sp^3d^2$, other] & one-hot \\
        aromaticity & 1 & aromatic or not [0/1] & one-hot \\
        hydrogens & 5 & number of connected hydrogens [0, 1, 2, 3, 4] & one-hot \\
        chirality & 1 & chiral or not [0/1] & one-hot \\
        \bottomrule
    \end{tabular}
    \caption{Description of atom features used in ReactAIvate}
    \label{tab:AtomFeatures}
\end{table*}

\begin{table*}[!ht]
    \centering
    \small
    \begin{tabular}{lrll}
        \toprule
        bond feature & size & description & feature type \\
        \midrule
        bond type & 4 & [single, double, triple, aromatic] & one-hot \\
        conjugation & 1 & conjugated or not [0/1] & one-hot \\
        ring & 1 & in ring or not & one-hot \\
        stereo & 4 & [StereoNone, StereoAny, StereoZ, StereoE] & one-hot \\
        \bottomrule
    \end{tabular}
    \caption{Description of atom features used in ReactAIvate}
    \label{tab:BondFeatures}
\end{table*}

\section{ReactAIvate} 

We proposed an interpretable graph attention model (ReactAIvate) for elementary step classification with concurrent identification of reactive atoms responsible for the chemical transformation of such steps. We followed the same protocol described in Attentive FP by \cite{xiong2019pushing}. The model was trained for 5 epochs with the batch size of 256. We have used accuracy(\%) as our performance metric. For the RAI task, we kept the default threshold value of 0.5. The performance with mean and standard deviation shown in Table 1 \& 2 of the main text is obtained using 5 different runs.

Model training involved the dataset containing 100,000 elementary steps, split into 70:10:20 ratios for training, validation, and testing, respectively. The test dataset is then used to get ID-RSC \& ID-RAI performance shown in Table 1 of the main text. Additionally, an out-of-distribution (OOD) dataset was created, comprising 3,647 elementary steps, with structurally distinct reaction components from those in the training set. This dataset is then used for OOD-RSC \& OOD-RAI tasks.


\subsection{Hyper-parameter tuning}

We conducted hyperparameter tuning for ReactAIvate, adjusting parameters such as $k$ (number of attentive message passing layers for atom embedding), $t$ (number of attentive message passing layers for molecule embedding), $h_{G}^*$ (dimension of graph feature size), and dropout ratio (d.r.). The results in Table \ref{tab:Hyperparameter} indicate that the accuracy for both tasks remains nearly unchanged by the hyperparameter adjustments.

\begin{table}[h]
\centering
\label{tab:Hyperparameter}
\begin{tabular}{lllllllllllll}
\toprule
 &&&&&ID-RSC&&ID-RAI&&OOD-RSC&&OOD-RAI&\\
Exp.&$k$&$t$&$h_{G}^*$&d.r.&acc.(\%)&F1&acc.(\%)&F1&acc.(\%)&F1&acc.(\%)&F1\\
\midrule
1&2&1&200&0.1&100.00&1.00&96.44&0.87&98.85&0.98&94.82&0.84\\
2&1&1&200&0.1&100.00&1.00&95.49&0.85&96.08&0.94&93.81&0.82\\
3&3&1&200&0.1&99.99&1.00&95.65&0.85&97.75&0.96&93.64&0.82\\
4&4&1&200&0.1&100.00&1.00&96.44&0.87&99.48&0.99&94.91&0.84\\
5&2&2&200&0.1&100.00&1.00&96.39&0.87&97.45&0.97&94.73&0.84\\
6&2&3&200&0.1&100.00&1.00&96.39&0.87&99.34&0.99&94.67&0.84\\
7&2&1&100&0.1&100.00&1.00&96.35&0.87&98.00&0.97&93.93&0.82\\
8&2&1&300&0.1&100.00&1.00&96.38&0.87&97.23&0.96&94.76&0.84\\
9&2&1&300&0.2&100.00&1.00&96.38&0.87&98.33&0.97&93.99&0.82\\
10&2&1&300&0.3&100.00&1.00&96.38&0.87&99.51&0.99&94.92&0.84\\
11&2&1&300&0.4&100.00&1.00&96.38&0.87&99.75&1.00&94.56&0.84\\

\bottomrule
\end{tabular}
\caption{Hyper-parameter tuning for ReactAIvate model}
\end{table}

\subsection{ReactAIvate with BCEloss}

We performed a control experiment using \texttt{BCELoss} instead of \texttt{WeightedBCELoss} corresponding to the RAI task, and the results are presented in Table \ref{table:table4}. In the OOD-RSC \& OOD-RAI tasks, ReactAIvate with \texttt{WeightedBCELoss} demonstrated superior performance compared to ReactAIvate with \texttt{BCELoss}.

\begin{table}[h]
    \centering
    \small
    \begin{tabular}{lcc}
        \toprule
        \textbf{Task} & \textbf{Accuracy} (\%) & \textbf{F1-Score}\\
        \midrule
        ID-RSC & 100.00 $\pm$ 0.00 & 1.00 $\pm$ 0.00\\
        ID-RAI & 97.52 $\pm$ 0.01 & 0.87 $\pm$ 0.01 \\
        OOD-RSC & 96.08 $\pm$ 4.2 & 0.92 $\pm$ 0.08\\
        OOD-RAI & 95.98 $\pm$ 0.01 & 0.84 $\pm$ 0.01\\
        \bottomrule
    \end{tabular}
    \caption{Performance of ReactAIvate with \texttt{BCELoss} for RSC and RAI tasks on ID and OOD test molecules}
    \label{table:table4}
\end{table}


\section{CRM generation using ReactAIvate}

\begin{figure*}[!ht]
	\begin{center}
		\begin{tabular}{c}
			\includegraphics[width=\columnwidth]{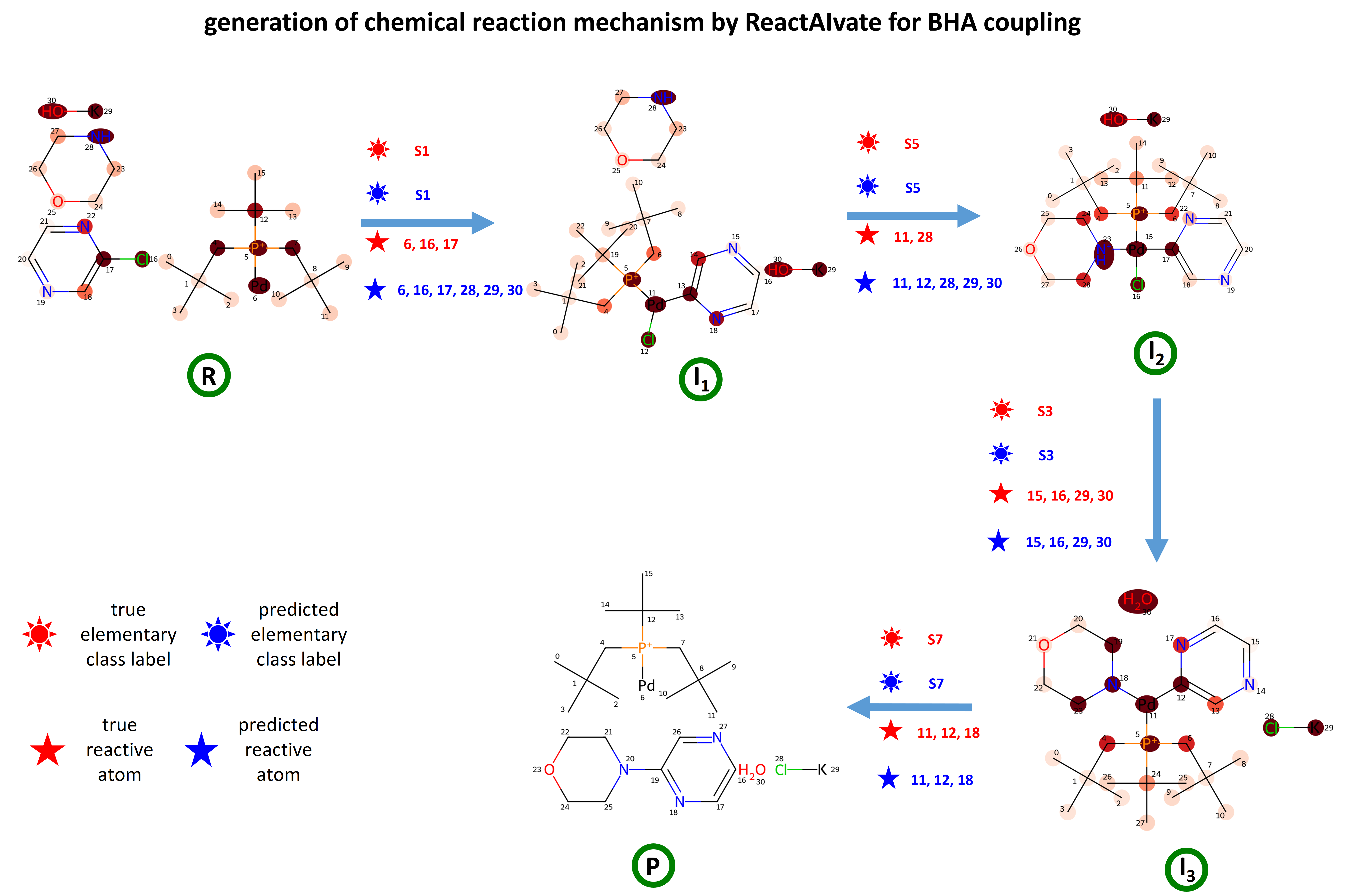}
		\end{tabular}
		\caption{BHA CRM generation using ReactAIvate. The model correctly predicted the sequence of elementary steps i.e., S1, S5, S3, and S7. The attentions are shown in each reactant/intermediate structure. The actual and predicted class labels as well as the reactive atoms are shown in each step.}
		\label{fig:fig6}
	\end{center}
\end{figure*}

In this section, we describe CRM generation using ReactAIvate. The first example is of a BHA reaction where a heteroaryl halide is coupled with a cyclic amine (morpholine) Figure \ref{fig:fig6}. The red colored eight-pointed star and five-pointed star shown above and below the arrows respectively denote a true elementary step in the mechanism and true reactive atoms respectively. The corresponding star notations in blue color are the predicted elementary step and reactive atoms respectively. The numbers that follow any of the five-pointed star notation are the atoms identified as important to the given elementary step of the reaction.

    The model predicted all the reaction classes accurately. In the first two steps (S1 and  S5) of the mechanism, a few more atoms are predicted as reactive atoms in addition to identifying the actual reactive atoms. For the last two steps (S3 and S7) all the reactive atoms are correctly predicted. It shall be noted that in the case of step S3, we have considered only four important atoms (Pd: 15, Cl: 16, K:29, and O:30) as the true reactive atoms. 

\begin{figure*}[!ht]
	\begin{center}
		\begin{tabular}{c}
			\includegraphics[width=\columnwidth]{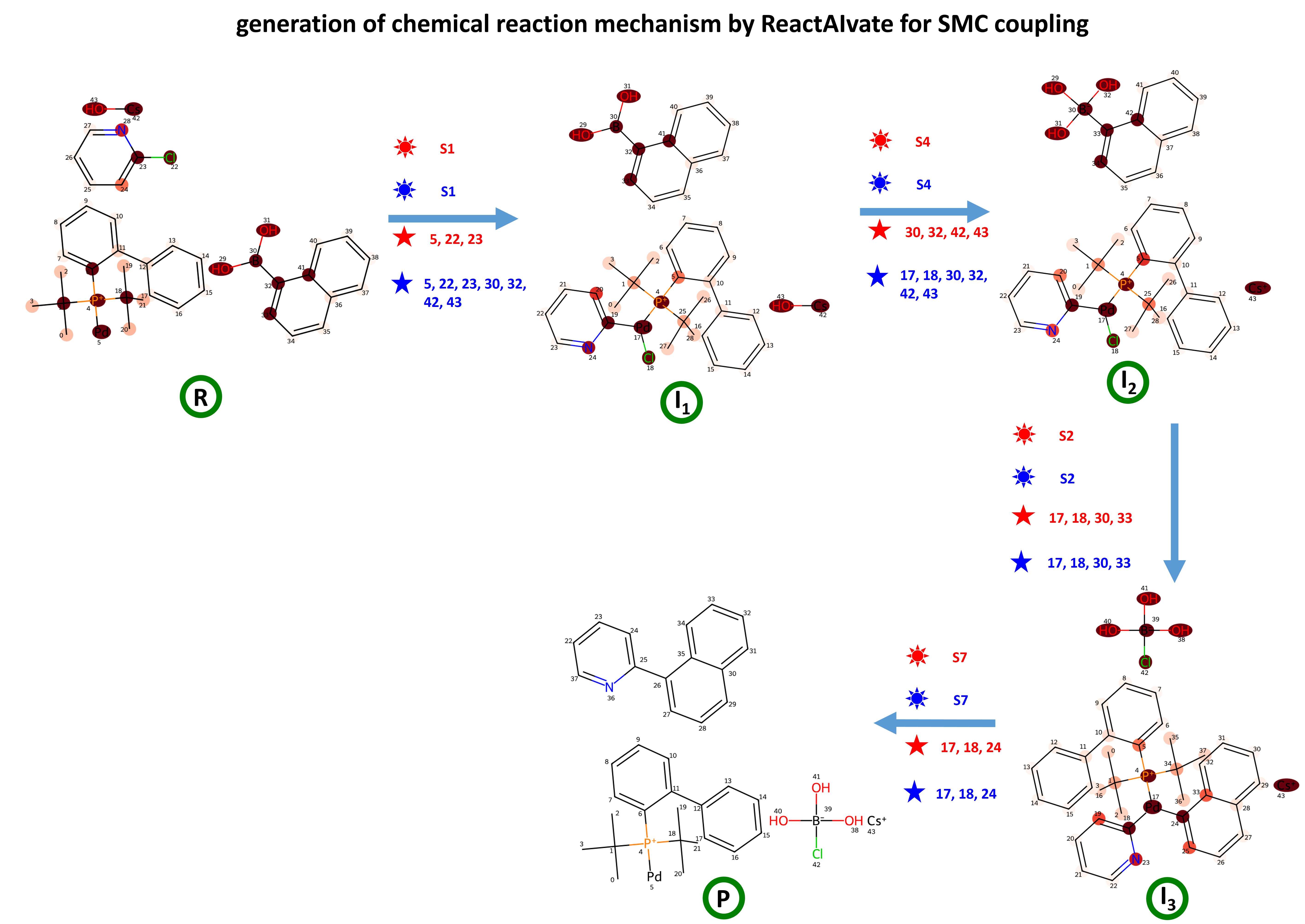}
		\end{tabular}
		\caption{An illustrative example of CRM generation of SMC reaction. The model correctly predicted the sequence of elementary steps i.e. S1, S4, S2, and S7. The attentions are shown in each reactant/intermediate structure. The actual and predicted class labels as well as reactive atoms are shown in each step.}
		\label{fig:fig7}
	\end{center}
\end{figure*}

\noindent The second example of CRM generation is shown in Figure \ref{fig:fig7} for a SMC reaction. Here, in contrast to the BHA reaction, a heteroaryl is coupled with an aryl boronic acid (naphthalene-1-boronic acid). Similar to the BHA CRM generation, all the elementary steps and the reactive atoms were predicted accurately. 

\section{Baseline models}

\subsection{T5Chem}

The T5Chem pipeline is a versatile language model designed for various chemical tasks, including forward reaction predictions \cite{lu2022unified}. The T5Chem is a pre-trained model trained on 97 million PubChem molecules. We have fined-tuned the T5Chem for our specific task of predicting the forward intermediate/product. The model takes a molecule in the form of SMILES strings. We used this model for forward synthesis such that given a set of reactants model learns to predict the corresponding CRM. T5Chem employs a complete encoder-decoder architecture featuring four layers and eight attention heads. The hidden dimension for T5Chem is set at 256, while the intermediate feed-forward layer utilizes a dimension of 2048. Starting with an initial learning rate of 5e-4, the model was trained over five epochs. The batch size was kept to 32. 
 
\subsection{Transformer}

We employed a simple transformer-based encoder-decoder architecture for CRM generation following the protocol described in \cite{rush2018annotated}. Both the encoder and decoder contains a stack of 4 identical layers of size 512 with 8 attention heads. We trained the model for 5 epochs and varied the learning rate using LambdaLR scheduler with ADAM optimizer and employed 3000 warm-up steps. The starting learning rate was set at 1.0. The batch size was kept at 64 and the sequences were padded up to 800 characters. Overall, the model had a total 29.5M parameters.

It is important to highlight that ReactAIvate has a total parameter count of only \emph{0.87M} in contrast to T5Chem with \emph{14.71M} and Transformer with \emph{29.55M} parameters. Thus, ReactAIvate is a simpler model as compared to the other two, which is also evident from the compute time required for model training, suggesting a higher efficiency.

\subsection{Deep Neural Network (DNN)}

We utilized a 3-layer DNN comprising of 512, 256, and 128 nodes respectively in the first, second, and third layer. Following each linear layer, ReLU activation function was employed, with a dropout rate of 0.1. Training spanned 100 epochs, using a learning rate of 0.0001 with the Adam optimizer, and a batch size of 128.

Various featurization methods were explored, including (1) the Morgan fingerprint with a radius of 2 and a bit vector length of 3096, (2) multiple fingerprint features with a bit vector length of 71208, and (3) RDKit-based 197 physicochemical descriptors. The parameter count for these three distinct models was 1.7M, 36.7M, and 0.3M, respectively.

\subsection{MPNN}

Our initial MPPN model is constructed using DGL-LifeSci and configured with the following hyperparameters. There are 39 input node features and 10 edge features. We applied 6 message passing steps during computation. Training was conducted for 5 epochs with a learning rate of 0.001, resulting in a model with 0.7M parameters.

\subsection{Limitations in CRM generation}

\begin{figure*}[!ht]
	\begin{center}
		\begin{tabular}{c}
			\includegraphics[width=0.9\columnwidth]{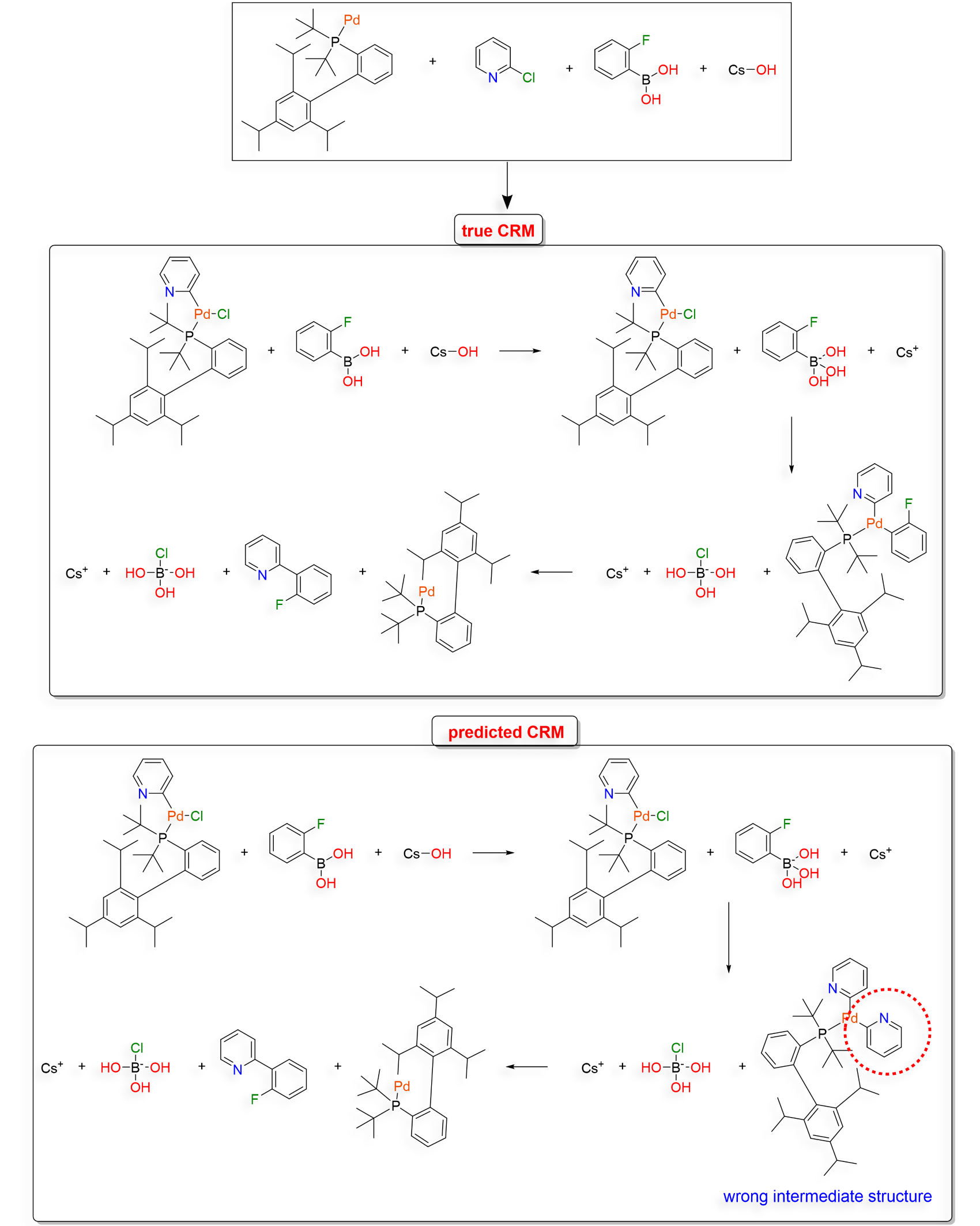}
		\end{tabular}
		\caption{Illustrative CRM generation for an in-distribution SMC sample using T5Chem model. The reacting entities are shown in the first box. The second and third boxes illustrate the true and the predicted mechanism respectively. The T5Chem model wrongly generated a pyridine ring instead of an o-fluoro aryl group, shown in red-dotted circles. See Figure \ref{fig:fig3} for more details of the SMC mechanism.}
		\label{fig:fig8}
	\end{center}
\end{figure*}

\begin{figure*}[!ht]
	\begin{center}
		\begin{tabular}{c}
			\includegraphics[width=0.9\columnwidth]{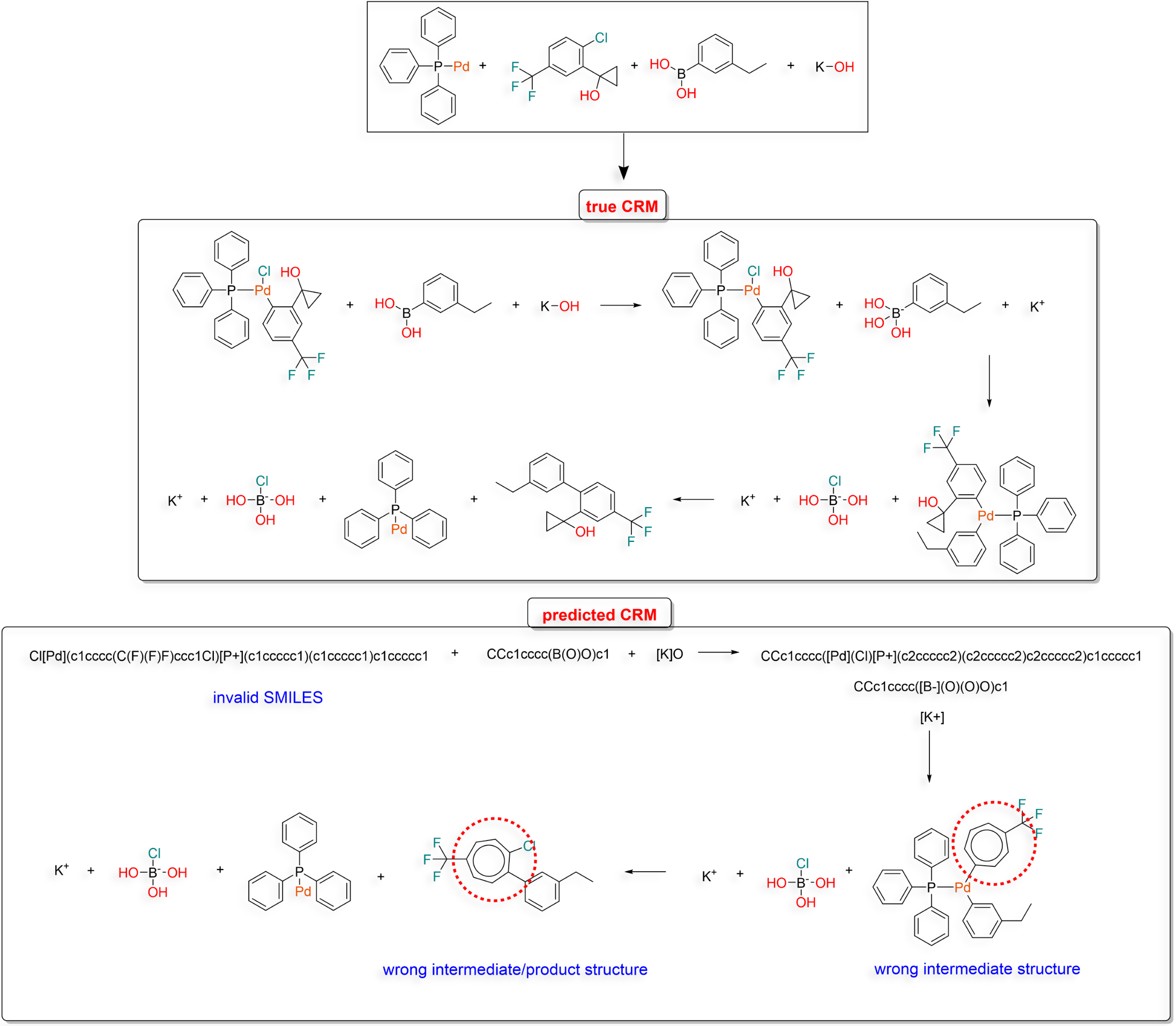}
		\end{tabular}
		\caption{An example of CRM generation for an out-of-distribution SMC sample using the T5Chem model. The reacting entities are shown in the first box. The second and third boxes respectively show the true and the predicted mechanism. It can be noted that in addition to incorrect molecule generation (shown in red-dotted circle), the model also generated invalid SMILES. For more details of the mechanism, see Figure \ref{fig:fig3}}
		\label{fig:fig9}
	\end{center}
\end{figure*}

\noindent Next, we have shown some examples of in-distribution (ID) and out-of-distribution (OOD) CRM generation tasks for both the baseline models.
The T5Chem offered a reasonably good accuracy in generating CRM for the in-distribution samples. However, it was found to result in the generation of semantically invalid  SMILES. For instance, in the predicted CRM Figure \ref{fig:fig8}, the intermediate generated after the boronate transmetallation step has a pyridine ring attached to the Pd center, instead of the actual o-fluoro aryl group. In the case of OOD CRM generation, the T5Chem model fails even in generating syntactically valid SMILES. Here also, semantically invalid molecules were generated (Figure \ref{fig:fig9}).

\begin{figure*}[!ht]
	\begin{center}
		\begin{tabular}{c}
			\includegraphics[width=0.8\columnwidth]{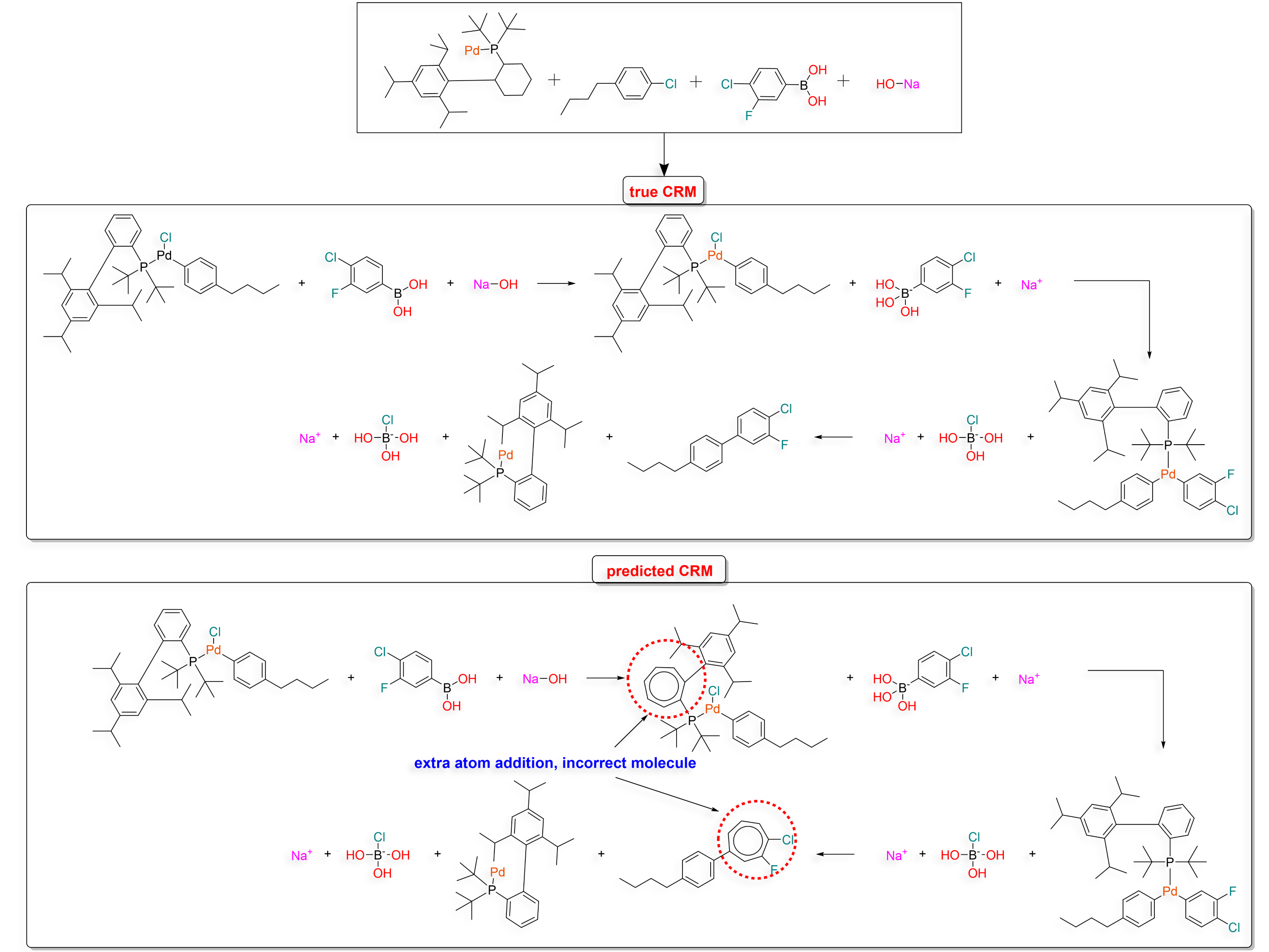}
		\end{tabular}
		\caption{An example of CRM generation for an in-distribution SMC sample using the Transformer model. The wrongly generated structures are shown in red-dotted circle. For more details of the mechanism, see Figure \ref{fig:fig3}}
		\label{fig:fig10}
	\end{center}
\end{figure*}

\begin{figure*}[!ht]
	\begin{center}
		\begin{tabular}{c}
			\includegraphics[width=0.8\columnwidth]{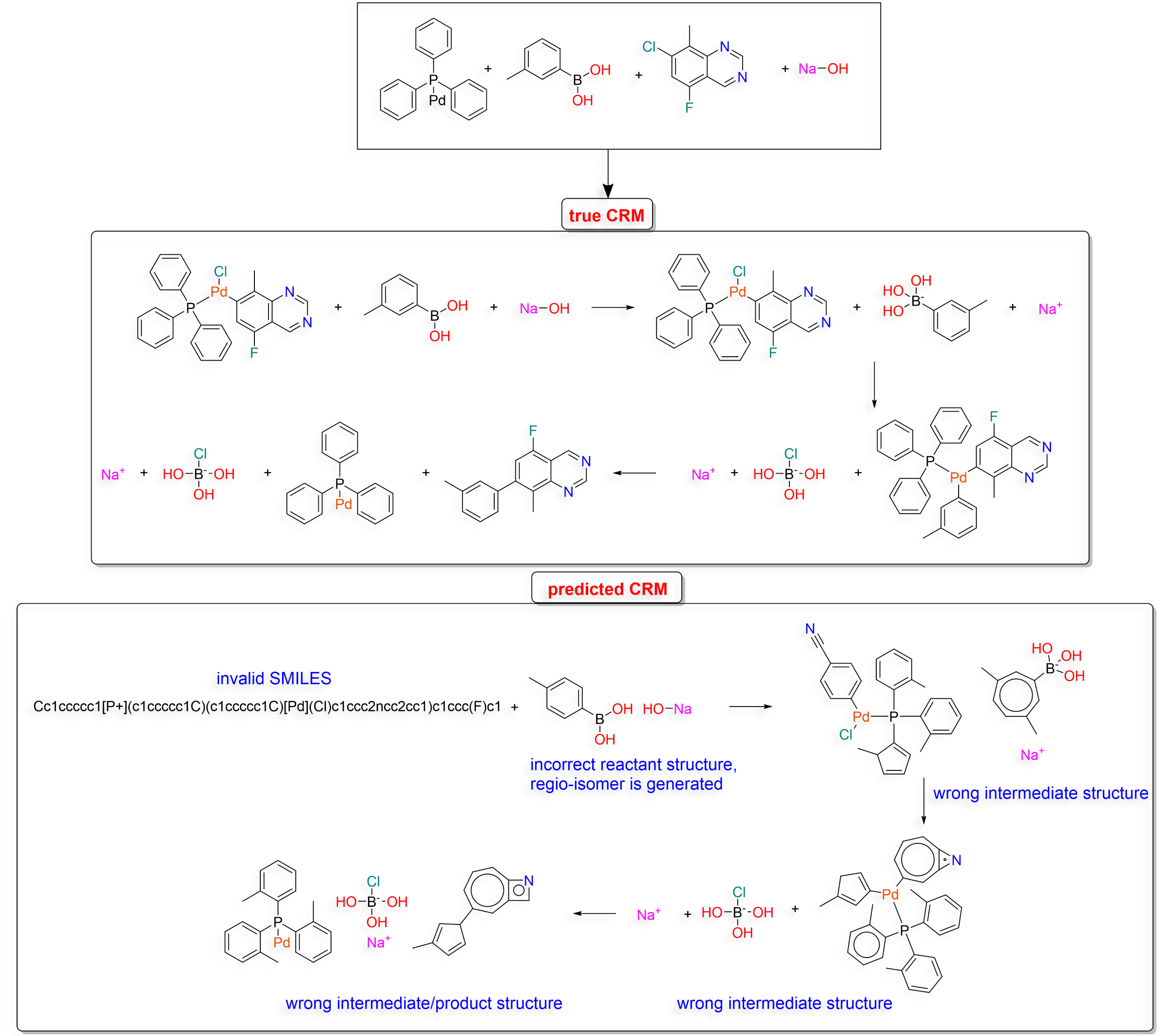}
		\end{tabular}
		\caption{An example of CRM generation for an out-of-distribution SMC sample obtained using the Transformer model. The wrongly generated structures are shown in red-dotted circles. The model also fails to generate semantically and syntactically valid SMILES. For more details of the mechanism, see Figure \ref{fig:fig3}}
		\label{fig:fig11}
	\end{center}
\end{figure*}

\noindent Similar to the T5Chem, the Transformer had issues arising from the generation of both syntactically and semantically invalid SMILES. See Figure \ref{fig:fig10} and Figure \ref{fig:fig11} respectively for Transformer generated CRM in ID and OOD samples respectively. 

\noindent Overall, we have shown that generating full CRM using the seq2seq model is problematic. Generation of semantically and syntactically invalid SMILES leads to invalid CRMs. 

\pagebreak

\pagebreak

\pagebreak



\bibliography{mybibfileSI}